\documentclass[aps,pra,twocolumn,superscriptaddress,groupedaddress]{revtex4}  

\usepackage{stmaryrd}
\usepackage[utf8]{inputenc}
\usepackage{graphicx}
\usepackage{color}
\usepackage{bm,bbold}
\usepackage{bbm}
\usepackage{enumerate}
\usepackage{color}
\graphicspath{figures} 
\usepackage{amsfonts, amsmath, amsthm, amssymb} 
\usepackage[colorlinks,bookmarks=false,citecolor=blue,linkcolor=red,urlcolor=blue]{hyperref}
\usepackage{changes}

\begin{document}

\title{Unveiling quantum entanglement in many-body systems from partial information}
\author{Irénée Frérot}
\email{irenee.frerot@neel.cnrs.fr}
\affiliation{ICFO - Institut de Ciencies Fotoniques, The Barcelona Institute of Science and Technology, 08860 Castelldefels (Barcelona), Spain}
\affiliation{Max-Planck-Institut f\"ur Quantenoptik, Hans-Kopfermann-Stra{\ss}e 1, 85748 Garching, Germany}
\affiliation{Univ Grenoble  Alpes, CNRS, Grenoble INP, Institut Néel, 38000 Grenoble, France}

\author{Flavio Baccari}
\email{flavio.baccari@mpq.mpg.de}
\affiliation{Max-Planck-Institut f\"ur Quantenoptik, Hans-Kopfermann-Stra{\ss}e 1, 85748 Garching, Germany}

\author{Antonio Acín}
\affiliation{ICFO - Institut de Ciencies Fotoniques, The Barcelona Institute of Science and Technology, 08860 Castelldefels (Barcelona), Spain}
\affiliation{ICREA - Institucio Catalana de Recerca i Estudis Avan\c cats, Lluis Companys 23, 08010 Barcelona, Spain}

\date{\today}

\begin{abstract}
Quantum entanglement is commonly assumed to be a central resource for quantum computing and quantum simulation. Nonetheless, the capability to detect it in many-body systems is severely limited by the absence of sufficiently scalable and flexible certification tools. This issue is particularly critical in situations where the structure of entanglement is \textit{a priori} unknown, and where one cannot rely on existing entanglement witnesses. Here, we implement a scheme in which the knowledge of the mean value of arbitrary observables can be used to probe multipartite entanglement in a scalable, certified and systematic manner. Specifically, we rely on positive semidefinite conditions, independent of partial-transposition-based criteria, necessarily obeyed if the data can be reproduced by a separable state. The violation of any of these conditions yields a specific entanglement witness, tailored to the data of interest, revealing the salient features of the data which are impossible to reproduce without entanglement. We validate this approach by probing theoretical many-body states of several hundreds of qubits relevant to existing experiments: a single-particle quench in a one-dimensional $XX$ chain; a many-body quench in a two-dimensional $XX$ model with $1/r^3$ interactions; and thermal equilibrium states of Heisenberg and transverse-field Ising chains. In all cases, these investigations have lead us to discover new entanglement witnesses, some of which could be characterized analytically, generalizing existing results in the literature. In summary, our paper introduces a flexible data-driven entanglement detection technique for uncharacterized quantum many-body states, of immediate relevance to experiments in a quantum advantage regime.
\end{abstract}

\maketitle

\section{Introduction}
\label{sec_introduction}
Quantum entanglement is a distinguished feature of composite quantum systems, marking a fundamental departure from their classical counterparts \cite{horodeckietal2009}. Over the last decade, it has become a commonplace that many-body entanglement represents an essential resource for quantum computation \cite{Preskill2018quantumcomputingin}, quantum simulation \cite{georgescuetal2014}, and quantum metrology \cite{pezzeetal2016}. While, on the theoretical side, the exact role of quantum entanglement in offering a quantum advantage remains somewhat controversial \footnote{
The necessity of entanglement to offer a quantum advantage in quantum metrology has been questioned \cite{braunetal2018}; concerning quantum simulation and computation, many efficient classical algorithms to simulate restricted families of entangled states have been developed, hence, entanglement is certainly not a sufficient ingredient to offer a quantum advantage. For pure states, however, it is necessary, as slightly entangled pure-state computations have an efficient classical representation~\cite{Vidal03}.}, the ability to manipulate quantum many-body superpositions arguably represents a major endeavour for many experimental platforms. As a matter of fact, the controlled preparation of many-body entangled states is a hallmark of such capability, and has been achieved in several experimental systems \cite{monzetal2011,islametal2015,kaufman_quantum_2016,songetal2017,wangetal2018,Friis_2018,omran_generation_2019,Brydges_2019,satzinger2021realizing}. On the other hand, a growing number of experiments operate in regimes inaccessible to the best available classical simulations \cite{Martinis2019,chiuetal2019,koepselletal2019,ebadi2020quantum,zhong_quantum_2020,scholl2020programmable,Bluvstein_2021}-- another hallmark pointing towards a genuine quantum advantage. It is commonly assumed that the intractability of classical simulations originates in the large-scale quantum entanglement which develops across the experimental system \cite{Preskill2018quantumcomputingin}. Nevertheless, a proper quantum computation, performed in a regime inaccessible to the best classical algorithms, and where the structure of quantum entanglement is also probed, has not yet been reported. This absence is partly due to the lack of sufficiently flexible and scalable theoretical tools to analyze the experimental data produced in such quantum devices. Surely, one cannot simply rely on the violation of existing entanglement witnesses, for the structure of entanglement in the system, and therefore the suitable entanglement criterion to potentially reveal it, are \textit{a priori} unknown. Furthermore, one cannot envision to use tomographic information about the underlying quantum state -- for acquiring such information would require a number of measurements growing exponentially with the system size \cite{paris2004quantum,Huang_2020}.

It is precisely the purpose of the present paper to { show the broad applicability of} a flexible and scalable tool to certify entanglement in an unknown multipartite quantum state.  In our setting, we assume that the expectation values of (a scalable number of) arbitrary observables are known.
Starting from the same insight as in \cite{bohnet_waldraff_etal_2017}, namely the connection between the compatibility of data with (quantum) separable states and the (classical) truncated moment problem, we provide a simple and scalable method to construct an entanglement witness from the observed expectation values that is tailored to be robust to noise.
The main insight from \cite{bohnet_waldraff_etal_2017} is that if the underlying quantum state is separable,  then the available data are obtained as entries of a correlation matrix, which satisfies certain positive-semidefinite constraints. 
Such compatibility conditions can be efficiently verified via so-called semidefinite-programming (SDP) techniques \cite{parillo_book}, allowing the study of systems of hundreds of qubits. The failure for the data to pass this test serves directly as an entanglement detection method, in which case our approach delivers a specific entanglement witness, violated by the observed data.

The resulting method is platform-agnostic, in the sense that how such data should be \textit{a priori} chosen, and how they should be inferred in an actual experiment is not relevant, and, in fact, will not be discussed in this work. As an illustration, we consider one- and two-body correlations for $N$ qubits, but our scheme is flexible and can incorporate the knowledge of any $k$-point function, or in general of any many-body observable.
By benchmarking the method on paradigmatic quench experiments, we show its wide applicability and its ability to extract physically relevant entanglement witnesses.
The expression of the witnesses themselves provides qualitative insight into the driving mechanism responsible for entanglement within the system. As a matter of fact, for several of the examples we have considered, we could analytically characterize the witnesses obtained numerically. This led us to extend some known entanglement criteria of the literature and derive completely new ones as well.  Analagously to the hierarchy introduced in \cite{bohnet_waldraff_etal_2017}, the scheme we propose can be generalised as a complete hierarchy of positive-semidefinite tests: if no separable state can reproduce the available data, the data will necessarily fail to pass the test at a finite level of the hierarchy -- in this sense, the hierarchy is complete.

\noindent\textbf{Comparison to previous works.} A large body of literature has already considered the problem of entanglement detection from partial information. Some of these results are recovered as special cases of the approach implemented in this paper; some others lack the scalability required to apply them to many-body systems; and some altenative scalable schemes either lack the flexibility of the present approach, or can be inconclusive. In particular, the so-called covariance matrix criterion \cite{gittsovichetal2010} and the generalized spin-squeezing inequalities \cite{tothetal2009}, which are based on one- and two-body correlations, are recovered as a consequence of our approach (as further discussed in Appendix \ref{app_previous_works}) -- while our approach is more flexible, as it can incorporate the knowledge of any correlation function. Criteria based on higher-order correlations have also been derived \cite{devincenteH2011,lietal2014,sarbickietal2020}. However, the efficiency of these approaches is unclear if only partial information is available (for example, if only two-body correlations are known). Furthermore, these approaches \cite{gittsovichetal2010,tothetal2009,devincenteH2011,lietal2014,sarbickietal2020} provide only sufficient conditions for entanglement, and therefore can be inconclusive even though the available data cannot be reproduced by a separable state; in contrast, here we provide a systematic and convergent hierarchy of criteria. A systematic approach, which can also incorporate partial knowledge about the quantum state, was proposed based on the solution of so-called separability eigenvalue equations \cite{sperlingetal2013,gerkeetal2018}; but this approach has an exponential cost and cannot be applied already to a few tens of qubits. Conceptually-different approaches, based on randomized measurements, have also been developed. Such approaches allow one to test bipartite entanglement criteria based on R{\'e}nyi entropies \cite{Brydges_2019,Huang_2020}, and PT-based criteria \cite{neven2021symmetryresolved,yu2021optimal}. However, in addition to the very high experimental requirements underlying these approaches, they require a number of measurements scaling exponentially with $N$, severely limiting their scalability beyond a few tens of qubits. Recently, intrinsically scalable approaches to the problem of multipartite entanglement detection from partial information have been developed. An entanglement-detection method from the knowledge of two-body reduced density matrices was developed in Ref.~\cite{navascues2020entanglement}, with a similar computational cost as the one in the present work; however, the above approach lacks the flexibility to be adapted to an arbitrary set of data, especially the average value of many-body observables. The approach of the present paper is complementary to Ref.~\cite{frerotR2021}, where the problem is solved through a mapping onto an inverse problem of classical statistical physics, offering a systematic and scalable solution; however this approach could be inconclusive for particular data sets. In contrast, here we solve a relaxation to this problem with an efficiency which is independent of the structure of the data, obtaining entanglement witnesses whose violation is guaranteed by semidefinite-positive constraints. Lastly, our approach shares some similarities with the method presented in~\cite{baccarietal2017}. However, in contrast to ours, the method in~\cite{baccarietal2017} is device-independent,  namely,  it exploits no information about the underlying Hilbert space.  This makes the resulting entanglement test sensitive to a careful choice of measurement basis for each particle.

The first entanglement detection approach based on the connection to the classical moment problem was introduced in Ref.~\cite{bohnet_waldraff_etal_2017}. What we develop here can be seen as a complementary separability test, based on the same conceptual premises, by with a different physical motivation.  While \cite{bohnet_waldraff_etal_2017} aims at constructing -- if it exists -- a separable state compatible with the data, our focus is instead on building a scalable and robust criterion for entanglement based on the available data.  Technically,  we define the SDP as a noise robustness problem, and restrict our analysis to the first level of a hierarchy of conditions in order to preserve scalability.  In contrast, the hierarchy in \cite{bohnet_waldraff_etal_2017} exploits a cost function tailored to identify a so-called ``flat extension'', which is a property that, by definition, can be assessed only by solving SDPs of increasing levels in the hierarchy. However, it should be emphasized that increasing by just one level the hierarchy is already extremely costly in a many-body setting, and it is not doable already for systems of few tens of particles.  Moreover, while an entanglement witness could potentially be obtained from the dual of the first level of the hierarchy in \cite{bohnet_waldraff_etal_2017}, it has no guarantee to be robust against noise.  Lastly,  we notice that the element of randomness in the objective function in \cite{bohnet_waldraff_etal_2017} implies that the witness will be different for every run of the SDP,  while our benchmarks allows one to derive analytical witnesses in many relevant scenarios.

In summary,  we introduced a systematic approach to multipartite entanglement detection in many-body systems from the knowledge of the average values of arbitrary observables, whose polynomial cost at every level is guaranteed with no assumptions about the structure of the data.  By benchmarking it on realistic many-body data, we are able to show that this approach has a wide range of applicability, and is able to recover and generalize several entanglement witnesses tailored to many-body systems of immediate experimental relevance. \\

In Section \ref{sec_framework}, we present our framework for data-driven entanglement detection. In Section \ref{sec_pair_of_qubits} we present an illustrative simple example for a Bell pair. In Section \ref{sec_results}, we apply our method to theoretical data of realistic many-body systems, both for quench experiments, and for thermal equilibrium states. Section \ref{sec_conclusion} displays our conclusions. More technical considerations on our method are given in Appendix \ref{sec_technical}, Appendix \ref{app_bipartite_witnesses} contains the detailed derivation of a new bipartite entanglement witness discovered through our approach, while Appendix \ref{app_previous_works} derives the entanglement criteria of refs.~\cite{gittsovichetal2010,tothetal2009} within our framework.

\section{Framework}
\label{sec_framework}

Some of the technical derivations of our entanglement detection method are similar to the approach presented in ref.~\cite{bohnet_waldraff_etal_2017}. For the sake of giving a comprehensive and self-contained description, we give complete introduction here, specialising it to the considered many-body setting.
We focus on a system composed of $N$ qubits (denoted $i \in \{1, 2, \dots, N\} =:[N]$), described by an unknown quantum state $\hat \rho$. We assume that the average values of several quantum observables $\hat{\cal O}_r$ are known. Our ultimate goal is to prove, only from the knowledge of these average values, that the quantum state $\hat \rho$ is entangled.This will be achieved by exhibiting a specific entanglement witness operator, in the form of a linear combination of the $\hat {\cal O}_r$ operators, which is violated by the data under consideration. These average values are either obtained by directly measuring the observables in question, or are inferred from other measurements \cite{Huang_2020}.  { Throughout this work, entanglement is defined as the impossibility to decompose the many-body density matrix as a statistical mixture of product states over individual qubits. This encompasses the situation where all qubits are individually addressed, as is the case in typical quantum computing or quantum simulation applications; but also the situation where the qubits are two-level subspaces of indistinguishable particles, as is often the case in atomic ensemble experiments. In this latter case, where the two levels can be either two spatial modes or two internal states, correlations among the qubits are only probed via collective measurements (typically, fluctuations of collective spin observables) \cite{pezzeetal2016}. Such information can be naturally incorporated in our approach in order to probe entanglement among the particles.} \\

\begin{figure}
\includegraphics[width=\linewidth]{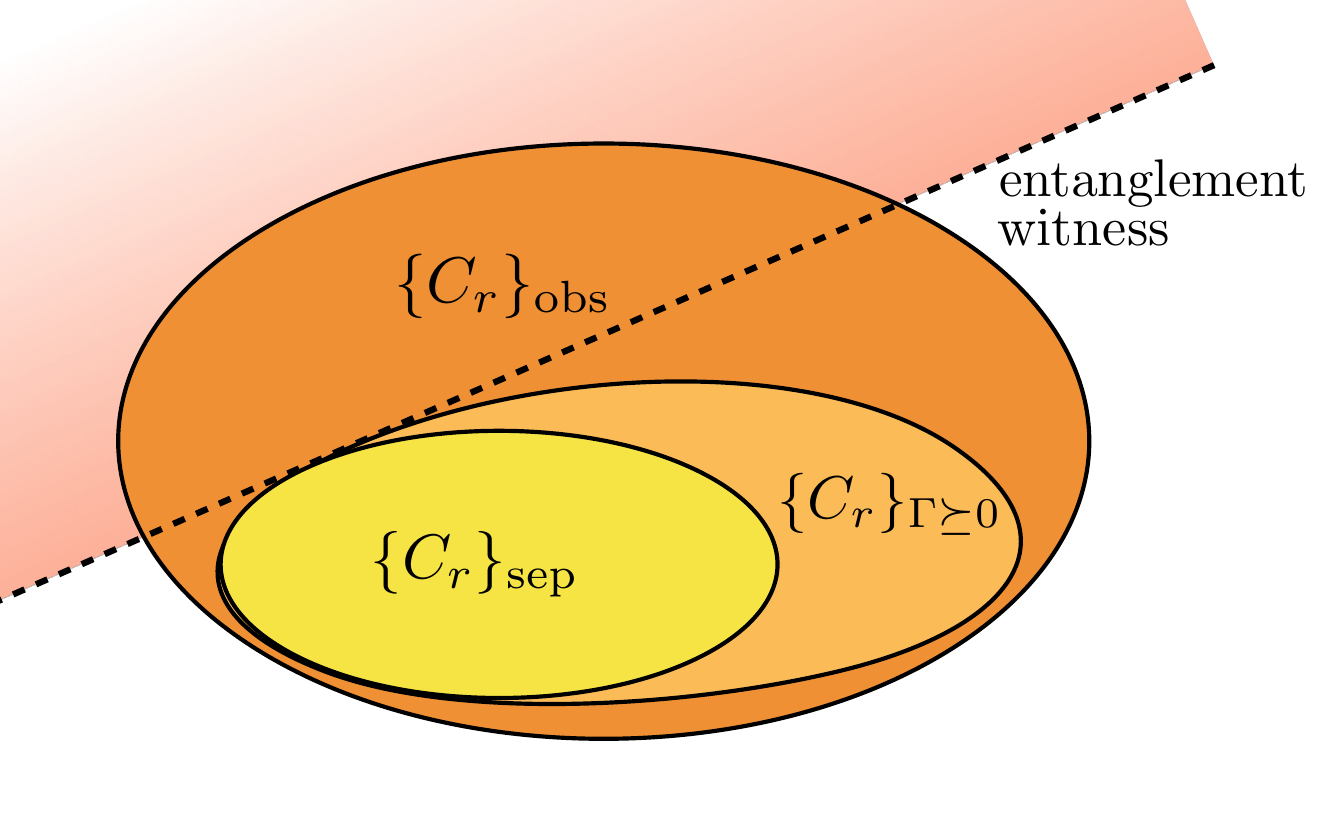}
\caption{Geometrical representation of the proposed entanglement detection framework. By arranging the observed data ${\cal D}_{\hat \rho}=\{C_r\}_{r=1}^R$ as a vector, one can represent them as a point in a $R$-dimensional space. Among all valid quantum data (represented as the dark orange convex set), one can identify the separable set, namely the convex subset of data which can be reproduced with a separable state (yellow set). We consider an efficient way to characterize a strict superset, corresponding to the data $\{C_r\}_{\Gamma \succeq 0}$ compatible with a positive-semidefinite correlation matrix $\Gamma$ (light orange set). Such a set contains $\lbrace C_r \rbrace_{\text{sep}}$; hence, if the data does not pass the $\Gamma \succeq 0$ test, then ${\cal D}_{\hat \rho}$ necessarily lies outside of the separable set, constituting a proof of entanglement. The method also provides an entanglement witness, i.e. a hyperplane separating the observed data from the separable set.     
}
\label{fig_Geometry}
\end{figure}

\noindent\textbf{Available quantum data.} For simplicity, in what follows we assume that some one- and two-body correlations have been obtained (the more general situation, where the average value of an arbitrary collection of operators is known, is discussed in Appendix \ref{sec_general_formulation}). We introduce the following notations for these data:
\begin{equation}
\label{eq_1body2body_correlators}
\left\lbrace
	\begin{array}{c}
	C_{i}^{X} = {\rm Tr}[\hat \rho \hat X_i] \\
	C_{i}^{Y} = {\rm Tr}[\hat \rho \hat Y_i] \\
	C_{i}^{Z} = {\rm Tr}[\hat \rho \hat Z_i]
	\end{array}
\right. ~~~~
\left\lbrace
	\begin{array}{c}
	C_{ij}^{XX} = {\rm Tr}[\hat \rho \hat X_i \hat X_j] \\
	C_{ij}^{XY} = {\rm Tr}[\hat \rho \hat X_i \hat Y_j] \\
	\vdots
	\end{array}
\right. ~,
\end{equation}
where $\hat X_i,\hat Y_i, \hat Z_i$ denote the qubit Pauli matrices. Notice that some of these correlators might be unknown. For instance, cross-terms such as $C_{ij}^{XY}$ or $C_{ij}^{XZ}$, whose measurement require individual addressing of the qubits, are often more challenging to infer than $C_{ij}^{XX},C_{ij}^{YY},C_{ij}^{ZZ}$ which can be measured via global rotations of all qubits before measuring in a fixed basis. We therefore allow for an incomplete data set ${\cal D}_{\hat \rho}=\{C_r\}_{r=1}^R$ composed of only a subset of all possible correlators [we introduce the generic notation $C_r := {\rm Tr}(\hat \rho \hat{\cal O}_r)$ to denote either $C_i^a$ or $C_{ij}^{ab}$ for some $1\le i < j \le N$; and some $a,b \in \{X,Y,Z\}$]. The method we develop verifies necessary conditions which are obeyed by ${\cal D}_{\hat\rho}$ if it can be reproduced by a separable state. The violation of any of these conditions leads our algorithm to produce a specific entanglement witness, tailored to the data under investigation, whose violation proves that the state $\hat \rho$ is entangled (see Fig.\ref{fig_Geometry} for a pictorial representation).\\ 

\noindent\textbf{Sufficient conditions for entanglement.}
By definition, a state is separable (i.e. not entangled) if it can be decomposed as a statistical mixture of product states:
\begin{equation}
	\hat \rho_{\rm sep} = \prod_{i=1}^N \int_{|{\bf n}_i|=1} p[\{{\bf n}_i\}] \otimes_{i=1}^N \hat \rho_{{\bf n}_i} ~.
	\label{eq_def_rho_sep}
\end{equation}
Here, we have represented the local states $\hat\rho_{{\bf n}_i}$ in a Bloch-sphere picture:
\begin{equation}
	\hat\rho_{{\bf n}_i} = |{\bf n}_i\rangle\langle{\bf n}_i|=\frac{1}{2}(\mathbb{1} + x_i \hat X_i + y_i \hat Y_i + z_i \hat Z_i) ~,
\end{equation}
where the local variables ${\bf n}_i = (x_i,y_i,z_i)$ satisfy:
 \begin{equation}
 	\forall i \in [N] ~~{\bf n}_i^2 = x_i^2 + y_i^2 + z_i^2 = 1 \label{eq_local_constraint_LV} ~.
 \end{equation}
It follows that $p[\{{\bf n}_i\}]\ge 0$ can be seen as a joint probability distribution over the unit (Bloch-sphere) vectors ${\bf n}_i$ for all the qubits $i = 1,\ldots,N$. Using that:
\begin{equation}
\label{def_local_variables} 
\left\lbrace
	\begin{array}{c}
	x_i = {\rm Tr}[\hat \rho_{{\bf n}_i} \hat X_i] \\
	y_i = {\rm Tr}[\hat \rho_{{\bf n}_i} \hat Y_i] \\
	z_i = {\rm Tr}[\hat \rho_{{\bf n}_i} \hat Z_i]
	\end{array}
\right. ~,
\end{equation}
one may express the correlators [Eq.~\eqref{eq_1body2body_correlators}] in a separable state [Eq.~\eqref{eq_def_rho_sep}] as classical expectation values over the $p$ distribution:
\begin{subequations}
\label{eq_1body_correlators}
\begin{align}
	&C_i^A = \int_{|{\bf n}_i|=1} p_i({\bf n}_i)~a_i =: \langle a_i \rangle \\
	&C_{ij}^{AB} = \int_{|{\bf n}_i|=1}\int_{|{\bf n}_j|=1} p_{ij}({\bf n}_i,{\bf n}_j)~a_ib_j =: \langle a_ib_j \rangle
	\end{align}
\end{subequations}
for $a,b \in \lbrace x,y,z\rbrace$. We denoted $p_i$ (resp.~$p_{ij}$) the marginal distribution over the $i$-th qubit (resp.~the $(i,j)$ pair); and introduced the notation $\langle \cdots \rangle$ for expectation values over the $p$ distribution.

In order to detect entanglement, one has to prove that the observed correlations $\{C_{i}^{a},C_{ij}^{ab}\}$ cannot be reproduced by the expressions \eqref{eq_1body_correlators} for any choice of joint probability distribution $p[\{{\bf n}_i\}]$.
Crucially, one can derive conditions which are necessarily satisfied if a distribution $p[\{{\bf n}_i\}]$ reproducing the data exists -- conditions whose violation is hence sufficient to conclude that the state $\hat \rho$ is entangled. In order to do so, one first defines the set of classical variables ${\bf m} = (1, x_1,y_1,z_1, \dots, x_N,y_N,z_N)$, and construct the correlation matrix $\Gamma_{\alpha,\beta} = \langle m_\alpha m_\beta \rangle$ over the $p$ distribution (more general choices of sets ${\bf m}$ might be considered, and are discussed in Appendix \ref{sec_general_formulation}). \newline 
The correlation matrix $\Gamma$
satisfies the following properties:

\begin{itemize}
    \item it is symmetric, $\Gamma = \Gamma^T$.
    \item it is positive semi-definite (PSD, i.e. $\Gamma \succeq 0$) by construction: indeed, for any vector ${\bf v}$, we have ${\bf v}^T \Gamma {\bf v} = \langle (\sum_\alpha m_\alpha v_\alpha)^2 \rangle \ge 0$.
    \item Some of its entries correspond to the observed data $C_i^a$ and $C_{ij}^{ab}$. For instance, for $m_{\alpha'} = x_i$ and $m_{\beta'} = x_j$, then $\Gamma_{\alpha',\beta'} = \langle x_i x_j \rangle$.
    \item Lastly, some remaining entries obey additional linear constraints, because of the condition \eqref{eq_local_constraint_LV}. In particular, we have $\langle z_i^2 \rangle = 1 - \langle x_i^2 \rangle - \langle y_i^2 \rangle$ for all $i$.
    
\end{itemize}

As an example, consider the case in which the one-body terms $C_i^X,C_i^Y,C_i^Z$ have been measured, together with two-body terms $C_{ij}^{XX},C_{ij}^{YY},C_{ij}^{ZZ}$. The corresponding $\Gamma$ reads:
\begin{widetext}
\begin{equation}
\Gamma = \Gamma^T =
\begin{pmatrix}
1 & {\color{blue}\bm{C_1^X}} & {\color{blue}\bm{C_1^Y}} & {\color{blue}\bm{C_1^Z}} & {\color{blue}\bm{C_2^X}} & {\color{blue}\bm{C_2^Y}} & \cdots & {\color{blue}\bm{C_N^X}} & {\color{blue}\bm{C_N^Y}} & {\color{blue}\bm{C_N^Z}} \\
\cdot & \langle x_1^2\rangle & \langle x_1y_1\rangle & \langle x_1z_1\rangle & {\color{blue}\bm{C_{12}^{XX}}} & \langle x_1y_2\rangle & \dots & {\color{blue}\bm{C_{1N}^{XX}}} & \langle x_1y_N \rangle& \langle x_1z_N\rangle \\
\cdot & \cdot & \langle y_1^2\rangle & \langle y_1z_1\rangle & \langle y_1x_2\rangle & {\color{blue}\bm{C_{12}^{YY}}} & \dots & \langle y_1x_N\rangle & {\color{blue}\bm{C_{1N}^{YY}}} & \langle y_1z_N\rangle \\
\cdot & \cdot & \cdot & 1-\langle x_1^2\rangle-\langle y_1^2\rangle & \langle z_1x_2\rangle & \langle z_1y_2\rangle & \dots & \langle z_1x_N \rangle& \langle z_1y_N\rangle & {\color{blue}\bm{C_{1N}^{ZZ}}}\\
\cdot & \cdot & \cdot & \cdot & \langle x_2^2\rangle & \langle x_2y_2\rangle & \dots & {\color{blue}\bm{C_{2N}^{XX}}} & \langle x_2y_N\rangle & \langle x_2z_N\rangle\\
\cdot & \cdot & \cdot & \cdot & \cdot & \langle y_2^2\rangle & \dots & \langle y_2x_N\rangle & {\color{blue}\bm{C_{2N}^{YY}}} & \langle y_2z_N\rangle \\
\cdot & \cdot & \cdot & \cdot & \cdot & \cdot & \ddots &  \vdots & \vdots & \vdots \\
\cdot & \cdot & \cdot & \cdot & \cdot & \cdot & \cdot & \langle{x_N^2}\rangle &  \langle{x_N y_N}\rangle&  \langle{x_N z_N}\rangle \\
\cdot & \cdot & \cdot & \cdot & \cdot & \cdot & \cdot & \cdot &  \langle{y_N^2}\rangle&  \langle{y_N z_N}\rangle \\
\cdot & \cdot & \cdot & \cdot & \cdot & \cdot & \cdot &  \cdot & \cdot &  1-\langle{x_N^2}\rangle-\langle{y_N^2}\rangle
\end{pmatrix} \succeq 0 
\label{eq_Gamma_matrix}
\end{equation}
\end{widetext}
Notice that we have marked in blue the entries replaced with the available data. All the other entries (black terms $\langle \cdots \rangle$) are unknowns which represent unobserved correlations over the $p$ distribution. If other correlators were known (e.g.~$C_{12}^{XY}$), they would simply replace the corresponding free variables in Eq.~\eqref{eq_Gamma_matrix} (namely $\langle x_1 y_2 \rangle$): reducing the number of free variables makes it harder to complete the matrix $\Gamma \succeq 0$, and therefore makes it easier to detect entanglement.
Notice that correlations such as $\langle x_i y_i \rangle$ have no experimental meaning in quantum physics, as they involve the simultaneous measurement of two incompatible observables, namely $\hat X_i$ and $\hat Y_i$ on the same qubit. However, they represent perfectly well-defined quantities if the state is separable, as (classical) expectation values over the $p$ distribution. Therefore, if the state is separable, it must be possible to complete the $\Gamma$ matrix with such unobserved correlations, such that $\Gamma \succeq 0$. Crucially, solving this problem is a so-called \textit{semi-definite program} \cite{parillo_book}, for which efficient convex-optimization algorithms are available. As we illustrate in Section \ref{sec_results}, the scalability of the method allows one to detect entanglement in systems of hundreds of qubits in a data-agnostic manner -- that is, without \textit{a priori} knowing the suitable entanglement criteria. \\

\noindent\textbf{Construction of an entanglement witness.} Importantly, if the matrix $\Gamma \succeq 0$ cannot be completed, the theory of semi-definite programming allows one to derive an entanglement witness of the form:
\begin{equation}
	\sum_{r=1}^{R} w_r C_r \stackrel{\hat \rho_{\rm sep}}{\le} 1 - \lambda \label{eq_witness}
\end{equation}
where the sum runs only over the available data. As discussed in Appendix \ref{sec_witness_qubits}, inequality \eqref{eq_witness} is satisfied by all separable states, while the data under investigation are such that:
\begin{equation}
	\sum_{r=1}^{R} w_r C_r = 1 ~, \label{eq_witness_violation}
\end{equation}
ultimately proving that the quantum state generating these data is entangled. The parameter $\lambda > 0$ in Eq.~\eqref{eq_witness} can be interpreted as the \textit{noise robustness} of the data. Indeed, if the quantum state $\hat \rho$ generating the data is mixed with white noise: $\hat \rho \to (1-\lambda)\hat \rho + \lambda \mathbb{1}/D$ with $D=2^N$ the dimension of the Hilbert space, then using the fact that Pauli observables are traceless, we have $\{C_i^a, C_{ij}^{ab}\} \to \{(1-\lambda)C_i^a, (1-\lambda)C_{ij}^{ab}\}$. The parameter $\lambda$ thus exactly quantifies the maximal amount of white noise which can be tolerated before entanglement detection becomes impossible with Eq.~\eqref{eq_Gamma_matrix}.  {  Hence,  by using an SDP to minimize the noise strength $\lambda$ for which the noisy data becomes compatible with a $\Gamma \succeq 0$, one obtains the maximally robust witness possible with the method} (see Appendix \ref{sec_witness_qubits} for details).
\newline

\noindent\textbf{A converging hierarchy of conditions.}
The presented approach can be extended to include also higher-order correlators. As further discussed in Appendix \ref{sec_general_formulation}, one may consider the set of classical variables ${\bf m}' = \{1\} \cup \{a_i\} \cup \{a_i b_j \} \cup \{a_i b_j c_k\} \cup \dots$, where $a_i, b_j, c_k$ are any components of the local classical variables $\{{\bf n}_i\}$. One then constructs the (PSD) correlation matrix $\Gamma_{\alpha,\beta}' = \langle m_\alpha' m'_\beta \rangle$ over the $p$ distribution. Verifying the compatibility of the data with $\Gamma'\succeq 0$ is again a semidefinite program, which can be solved at a computational (memory) cost scaling at most as ${\cal O}({\rm length}({\bf m}')^2)$. The matrix $\Gamma$ [Eq.~\eqref{eq_Gamma_matrix}] is obtained as a submatrix of $\Gamma'$, and therefore the condition $\Gamma' \succeq 0$ is stronger than $\Gamma \succeq 0$. Including in ${\bf m}'$ all monomials up to degree $l=1,2,3,\dots$, one defines a systematic hierarchy of positive-semidefinite conditions which are necessarily obeyed if the underlying state $\hat \rho$ is separable. Crucially, as further discussed in Appendix \ref{sec_general_formulation}, if no separable state can reproduce the data, then there exists a finite degree $l$ such that the data fail to fulfill the corresponding condition $\Gamma'\succeq 0$ -- this property is a consequence of the variables $\{{\bf n}_i\}$ being compact: $|{\bf n}_i|=1$ for all $i$. Therefore, the approach presented here defines, in the limit $l \to \infty$, a converging hierarchy of outer approximations to the set of separable states, exhausting the capability of a given data set to demonstrate multipartite entanglement. The computational cost ${\cal O}((3N)^{2l})$ is strictly polynomial at each relaxation level. One may regard such a hierarchy as an instance of Lassere's relaxation of the moment problem for the probability distribution $p[\{{\bf n}_i\}]$ \cite{lasserre2001global,parillo_book}. Notice that in practice, the computational cost of higher-level tests ($l\ge 2$) increases rapidly, especially for hundreds of qubits. However, we provide in Section \ref{sec_results} compelling evidence of the efficiency and tightness of Eq.~\eqref{eq_Gamma_matrix}, which represents the $l=1$-level of the hierarchy, to detect entanglement in many-body systems in a flexible, unbiased and scalable manner.\\
 
\noindent\textbf{Invariance under partial transposition.} It is interesting to notice that our criteria are independent of the partial-transposition (PT) criteria \cite{Peres1996,dohertyetal2005,neven2021symmetryresolved,yu2021optimal}: a state $\hat \rho$ is compatible with Eq.~\eqref{eq_Gamma_matrix} if and only if (iff) the state $\hat \rho^{\rm PT}$ is compatible with Eq.~\eqref{eq_Gamma_matrix}, where $\hat \rho^{\rm PT}$ is obtained by applying partial transposition on any subset of qubits. Indeed, PT leaves invariant the Pauli matrices $\hat X_i$ and $\hat Z_i$, while $\hat Y_i$ is changed into $-\hat Y_i$. Therefore, all correlations involving $\hat Y_i$ are changed into their opposite. The corresponding matrix $\Gamma^{\rm PT}$ is then obtained from $\Gamma$ [Eq.~\eqref{eq_Gamma_matrix}] by a simple change of basis, in which $(x_i, y_i, z_i)$ is changed into $(x_i, -y_i, z_i)$ for the qubits where PT is applied. Therefore, $\Gamma^{\rm PT}$ can be completed as a PSD matrix iff $\Gamma$ can be completed as a PSD matrix. As discussed in Section \ref{sec_general_formulation}, this simple observation can be extended to the complete hierarchy of criteria derived via our approach.

\section{Simple two-qubit example}
\label{sec_pair_of_qubits}

As a first illustration of the method, we consider an isotropic Werner state \cite{werner1989}, namely a statistical mixture of white noise with a spin singlet:
\begin{subequations}
\label{eq_def_werner_state}
\begin{align}
    &\hat \rho_\lambda = (1-\lambda) |\Psi\rangle\langle\Psi| + \frac{\lambda}{4}\mathbb{1} ~,\\
	&|\Psi\rangle = \frac{1}{\sqrt{2}}(|\uparrow \downarrow\rangle - |\downarrow\uparrow\rangle)
\end{align}
\end{subequations}
where $0\le \lambda \le 1$. The state $\hat \rho_\lambda$ is separable iff $\lambda \ge 2/3$. Let us show that Eq.~\eqref{eq_Gamma_matrix} is tight for the Werner state, namely that it detects entanglement whenever $\lambda<2/3$. The Werner state is $SU(2)$ invariant, and one finds $C_1^a=C_2^a=0$ (for $a \in \{X, Y, Z\}$) and $C_{12}^{ab}=-\delta_{ab}(1-\lambda)$. However, this detailed property, impossibly to exactly fulfill in an experiment, is not needed to demonstrate entanglement with our method. It turns out to be sufficient to consider only $c:=C_{12}^{XX} + C_{12}^{YY} + C_{12}^{ZZ}$ as available data. As discussed in Appendix \ref{app_symmetries}, if only $c$ is known and without making any assumption about the underlying quantum state, one may symmetrize the distribution $p(\{x_i, y_i, z_i\})$ aimed at reproducing the data with a separable state, Eq.~\eqref{eq_def_rho_sep}. This leads us to drastically simplify Eq.~\eqref{eq_Gamma_matrix} as:
\begin{equation}
	\Gamma = \begin{pmatrix}
		1 & 0 & 0 & 0 & 0 & 0 & 0 \\
		\cdot & 1/3 & 0 & 0 & {\color{blue} \bm{c/3}} & 0 & 0\\
		\cdot & \cdot & 1/3 & 0 & 0 & {\color{blue} \bm{c/3}} & 0\\
		\cdot & \cdot & \cdot & 1/3 & 0 & 0 & {\color{blue} \bm{c/3}}\\
		\cdot & \cdot & \cdot & \cdot & 1/3 & 0 & 0\\
		\cdot & \cdot & \cdot & \cdot & \cdot & 1/3 & 0 \\
		\cdot & \cdot & \cdot & \cdot & \cdot & \cdot & 1/3
	\end{pmatrix} \succeq 0
\end{equation}
Reorganizing the lines and columns, the matrix $\Gamma$ is block diagonal, and is PSD iff all blocks are PSD, that is, iff $
	\begin{pmatrix}
		1 & c \\
		c & 1
	\end{pmatrix} \succeq 0$,
iff $|c| \le 1$. This establishes that entanglement is detected whenever $|C_{12}^{XX} + C_{12}^{YY} + C_{12}^{ZZ}| > 1$, which is a (tight) entanglement witness already well known in the literature, and which is recovered by our approach. The witness is tight, as the product state $|\uparrow\uparrow\rangle$ is s.t. $C_{12}^{XX} + C_{12}^{YY} + C_{12}^{ZZ}= C_{12}^{ZZ} = 1$. In the case of the Werner state, we have $c=-3(1-\lambda)$, from which we recover the known result that the state \eqref{eq_def_werner_state} is entangled for $\lambda < 2/3$. Notice that Eq.~\eqref{eq_Gamma_matrix} is not tight for all two-qubit states: by producing random two-qubit states, we could find entangled states (as detected by the concurrence criterion \cite{wooters1998}) which are nevertheless compatible with a PSD correlation matrix as in Eq.~\eqref{eq_Gamma_matrix}.

\section{Robust detection of entanglement in many-body systems}
\label{sec_results}
We have then chosen to benchmark our entanglement detection method on paradigmatic lattice quantum spin models. Motivated by an ultracold atoms experiment \cite{gross_quantum_2017} realized a few years ago \cite{fukuharaetal2015}, we have first focused on the entanglement generated by a single impurity propagating along a one-dimensional $XX$ chain (Section \ref{sec_fukuhara}). As a main result, we found that the robustness of entanglement detection can be increased by about one order of magnitude using our method as compared to existing criteria, using the same data as collected in the experiment of Ref.~\cite{fukuharaetal2015}. We have then considered a two-dimensional system, where entanglement is generated by a $XX$ Hamiltonian with $1/r^3$ interactions, from an initial state with all spin polarized along $X$ (Section \ref{sec_2dXX}). This example of a many-body quench with power-law interactions is especially motivated by Rydberg arrays \cite{Levine_2018,Lienhard_2018,Keesling_2019,deleselucetal2019,browaeys_lahaye_2020,
scholl2020programmable,ebadi2020quantum,Bluvstein_2021}, ultracold magnetic atoms \cite{Fersterer_2019,Patscheider_2020,Gabardos_2020}, nitrogen-vacancy centers in diamond \cite{doherty_2013,Choi_2019} and trapped ions systems \cite{Jurcevic_2014,Richerme_2014,zhang_observation_2017,Friis_2018,Brydges_2019,Landsman_2019,Joshi_2020,Monroe_2021}, where related spin Hamiltonians have been implemented. In this case, our algorithm lead us to discover a wide family of entanglement witnesses based on components of the structure factor, which extend similar criteria reported previously in the literature, and which are especially suited to detect entanglement in out-of-equilibrium situations. Finally, we have explored the possibility to detect bipartite entanglement with our method, focusing on thermal-equilibrium states of the Heisenberg model and of the transverse-field Ising model in one dimension (Section \ref{sec_bipartite_witnesses}). Overall, these examples validate our method as a robust, flexible, effective and highly scalable approach to detect multipartite entanglement from partial information, as is currently collected in intermediate-scale quantum simulators and computers. 

{  Importantly,  since our main objective is to derive witnesses that can tolerate a realistic amount of noise,  we always solve the entanglement SDP test as a noise robustness problem.  That is, we mix the considered quantum states $\rho$ with white noise,  modelled as a completely mixed state: $\hat\rho \to (1-\lambda)\hat\rho + \lambda \mathbb{1}/2^{N}$.  The noise robustness $\lambda^*$ is then defined as the value of $\lambda$ above which entanglement is not detected any more by the moment matrix criterion.  By doing so,  the witness obtained by the dual of the SDP tolerates, by construction,  at least the amount of noise $\lambda^*$ (cf.  Sec.  \ref{sec_framework} and Appendix \ref{sec_witness_qubits} for details). }
To generate the numerical SDP problems, we use the software Ncpol2sdpa~\cite{algorithm2015wittek}, and we solve the SDPs with Mosek~\footnote{Available at \url{http://www.mosek.com/}}. We release an open source code \footnote{\url{https://github.com/ifrerot/SDP_multipartite_entanglement}}, which allows to recover results of Section \ref{sec_fukuhara}, and can be adapted to probe more general data.\\

\subsection{Single-spin-flip in a one-dimensional chain}
\label{sec_fukuhara}

\begin{figure*}
\includegraphics[width=\linewidth]{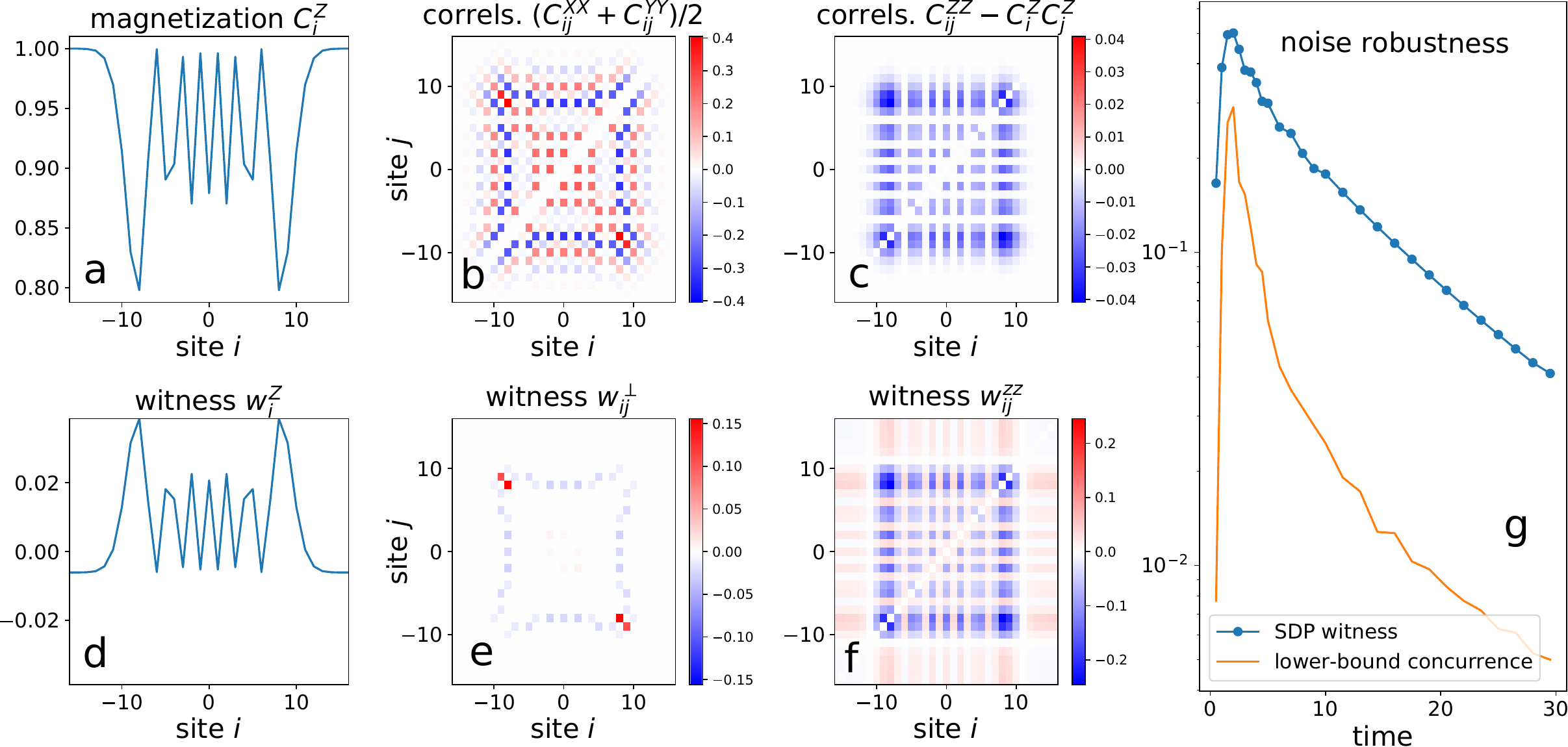}
\caption{Single spin flip in a $XX$ chain with $N=64$ spins (cf.~experiment of Ref.~\cite{fukuharaetal2015}). Panels (a,b,c): one- and two-body correlations at time $tJ=10$, used as input to our SDP algorithm (for the sake of readability, on panel c we plot the connected correlations $C_{ij}^{ZZ} - C_i^Z C_j^Z$). Panels (d,e,f): corresponding coefficients in the optimal entanglement witness. Panel (g): on a semilogarithmic scale, noise robustness of the violation of the witness as a function of time (blue dots); and theoretical prediction for the noise robustness of the concurrence lower-bound measured in \cite{fukuharaetal2015} (solid orange line).
}
\label{fig_witness_fukuhara}
\end{figure*}

We consider a one-dimensional ferromagnetic $XX$ chain with nearest-neighbour interactions:
\begin{equation}
	\hat {\cal H}_{XX} = -J \sum_{i=0}^{N-1} [\hat X_i \hat X_{i+1} + \hat Y_i \hat Y_{i+1}] ~,
\end{equation}
with $J=1$ a global energy scale, and with periodic boundary conditions.
As initial state, we consider the ferromagnetic state $|\Psi_0\rangle = \otimes_i |\uparrow \rangle$. We assume that at time $t=0$, the spin at $i=0$ is flipped into $|\downarrow\rangle$ \cite{Mazza_2015,fukuharaetal2015}. This central excitation then propagates along the chain under the XX Hamiltonian. As $\hat{\cal H}_{XX}$ conserves the total magnetization along $Z$, the dynamics occurs in the $N$-dimensional manifold of states generated by $\{\hat \sigma_i^- |\Psi_0\rangle\}_{i=0}^{N-1}$ [with $\hat \sigma_i^- = (\hat X_i - i \hat Y_i)/2$ the lowering operator]. Even though this simple quench is in essence a single-particle problem, multipartite entanglement is generated across the entire system. In the experiment of Ref.~\cite{fukuharaetal2015}, the propagation of entanglement was observed through a lower-bound to the pairwise concurrence \cite{Mazza_2015}, which measures the entanglement of the two-body reduced state $\hat \rho_{ij}$ \cite{wooters1998}. Here, our main result is that using the same information as in the experiment of Ref.~\cite{fukuharaetal2015} [namely, the transverse correlations $C_{ij}^\perp := (C_{ij}^{XX} + C_{ij}^{YY})/2$, the magnetization $C_i^Z$ and the longitudinal correlations $C_{ij}^{ZZ}$], more robust detection of entanglement is possible thanks to our method. 

In order to theoretically compute the spin-spin correlations, we assume periodic boundary conditions on a chain of $N=64$ spins (these choices have no visible effect on the results if the time is not long enough for the excitation to travel across the whole chain). This leads to:
\begin{subequations}
\label{eq_data_Fukuhara_quench}
\begin{align}
&\varphi_r(t) = N^{-1}\sum_{k=0}^{N-1}\exp\left[\frac{2i\pi kr}{N} + it\cos\left(\frac{2\pi k}{N}\right)\right] \\
&	C_i^Z = 1 - 2|\varphi_i|^2 \\
&	C_{ij}^{\perp} = 2\Re(\varphi_i^* \varphi_j) \\
&	C_{ij}^{ZZ} = 1 - 2(|\varphi_i|^2 + |\varphi_j|^2)
\end{align}
\end{subequations}
As discussed in Appendix \ref{app_symmetries}, in order to implement our algorithm, we may use the symmetries of the problem to drastically reduce the number of non-zero variables in Eq.~\eqref{eq_Gamma_matrix}, greatly improving the scalability. The resulting witness at each time, reconstructed via the algorithm described in Appendix \ref{sec_witness_qubits}, is then tailored to the structure of correlations at that particular time, and follows the propagation of the excitation along the chain. The witness operator is of the form $\hat W = \sum_i w_i^Z \hat Z_i + \sum_{i\neq j} [w_{ij}^{ZZ}\hat Z_i \hat Z_j + w_{ij}^{\perp}(\hat X_i \hat X_j + \hat Y_i \hat Y_j)/2]$.

In Fig.~\ref{fig_witness_fukuhara}, we plot for time $tJ=10$ the correlations used as input to the SDP algorithm (upper row), and the coefficients of the corresponding entanglement witness (lower row). { Both our witness and the concurrence lower-bound maximized over all pairs use the exact same data to detect entanglement.  In order to compare their respective strength in a meaningful way,  we have chosen to compute the noise robustness of the concurrence lower-bound as well.  In Fig.~\ref{fig_witness_fukuhara}(g), we plot the evolution as a function of time of the noise robustness for both our witness, and for} the concurrence lower-bound \cite{Mazza_2015} as measured in the experiment of Ref.~\cite{fukuharaetal2015}. The SDP witness is about one order of magnitude more robust againt white noise than the concurrence lower-bound. Clearly, beyond the quantitative information provided by the noise robustness, the structure of the witness also provides qualitative insight into the distribution of multipartite entanglement across the system. In particular, as is apparent in the transverse coefficients $w_{ij}^\perp$ [Fig.~\ref{fig_witness_fukuhara}(e)], the qubits whose contribution to the witness is the largest are located close to $i=-j=\pm vt$ (with $v=1$ the group velocity of the excitation). This feature is also captured by the two-body concurrence which is maximal for this pair of qubits. However, the precise contribution of other correlations is crucial to obtain a robust entanglement witness, as established by our data-agnostic approach. Finally, we notice that the separable bound as obtained from the SDP is tight, as we could always saturate this bound by a variational search over product states.

The single-particle nature of this problem is reflected in the fact that multipartite entanglement is progressively diluted throughout the system while the excitation, initially localized at $i=0$, spreads across the whole chain. As a consequence, at long times, the robustness of the violation decreases to zero for large systems [Fig.~\ref{fig_witness_fukuhara}(g)]. In the following example, we consider instead a genuine many-body problem where the entanglement generated by the unitary dynamics is robust at all times.

\subsection{Many-body quench dynamics in a two-dimensional power-law XX model}
\label{sec_2dXX}
We now consider a two-dimensional $XX$ model with $1/r^3$ interactions:
\begin{equation}
	\hat {\cal H}_{XX} = J \sum_{1\le i < j \le N} \frac{\hat X_i \hat X_j + \hat Y_i \hat Y_j}{r_{ij}^3} + h\sum_{i=1}^N \hat X_i ~,
	\label{eq_H_XXLR_2d}
\end{equation}
where $r_{ij}$ denotes the distance between spins $i$ and $j$, arranged over a $N = L \times L$ square lattice. We consider both open- and periodic boundary conditions with $N=400$ spins. We set $J=1$; and the transverse field $h=0.5$ is introduced for technical reasons (see below). This model is of direct relevance both to Rydberg arrays \cite{browaeys_lahaye_2020}, and to trapped ions \cite{Monroe_2021}. As initial state, we consider a ferromagnetic state along $X$: $|\Psi_0\rangle = \otimes_i |+\rangle_i$ with $|+\rangle = (|\uparrow\rangle + |\downarrow\rangle)/\sqrt{2}$. For this particular initial state, the dynamics is invariant under the change $\hat {\cal H}_{XX} \to -\hat {\cal H}_{XX}$; and $|\Psi_0\rangle$ represents the mean-field ground state of $-\hat {\cal H}_{XX}$ \cite{XXZLR}. The dynamics is then well-approximated by a semiclassical spin-wave approach \cite{multispeed}, involving bosonic gaussian states, whose stability is further enhanced by introducing the symmetry-breaking term $h\sum_i \hat X_i$. We would like to emphasize that simulating the exact dynamics of a quantum many-body system is a central issue for all numerical approaches, and we selected this particular example, amenable to a semiclassical treatment, for the sake of illustrating the suitability of our entanglement-detection method to large-scale systems with no translation invariance, as investigated in existing experimental platforms. Ultimately, our method unveils the (in)compatibility of a given set of correlations with a separable state, and the way in which these correlations were obtained (through exact computation, using some approximations as we achieve here through a spin-wave approach, or experimentally) is totally irrelevant to the method itself.
As input data, we have used the one-body terms $C_i^X$ (for all qubits $i$) and the two-body terms $C_{ij}^{aa}$ (for $a=X,Y,Z$ and all pairs $i<j$). Once again, we used symmetries to reduce the number of free variables in the SDP algorithm (see Appendix \ref{app_symmetries}, and Appendix \ref{sec_witness_qubits} for details on the algorithm used to reconstruct an entanglement witness from the data). We have first considered systems with periodic boundary conditions, such that the data are translationally invariant (TI). In this case, we could analyze analytically the witnesses found by our algorithm, and generalize them to a whole family of entanglement witnesses in its own right. We have then considered systems with open boundary conditions, illustrating the scalability of our approach to detect entanglement in generic (non-TI) systems with hundreds of qubits in a data-agnostic manner.\\

\noindent\textbf{A family of entanglement witnesses.}
Investigating TI systems, and generalizing the entanglement witnesses reconstructed by our algorithm, we found the following family of witnesses:
\begin{equation}
	\sum_{a\in\{X,Y,Z\}} \sum_{j \neq j'} e^{i[\phi_a(j') - \phi_a(j)]} C_{jj'}^{aa} \ge -N ~,
	\label{eq_TI_witness}
\end{equation}
where $\phi_a(j)$ are arbitrary local phases, potentially depending on the spin direction $a$. The proof of the separable bound is straightforward: assuming that the state is separable, we may introduce the local variables $x_i, y_i, z_i$ parametrizing the local quantum states (Section \ref{sec_framework}). Using that $N = \sum_i (x_i^2 + y_i^2 + z_i^2)$, we then have: $\sum_{a\in\{X,Y,Z\}} \sum_{j \neq j'} e^{i[\phi_a(j') - \phi_a(j)]} C_{jj'}^{aa} + N \stackrel{\rm sep}{=} \sum_{a \in\{X, Y, Z\}} \langle |\sum_j e^{-i\phi_a(j)} a_j|^2 \rangle \ge 0$. 

The witness of Eq.~\eqref{eq_TI_witness} turns out to be very similar to existing results in the literature \cite{krammeretal2009,crameretal2011,haukeetal2013}. There are however two important differences: on the technical side, the witness of Eq.~\eqref{eq_TI_witness} involves local phases $\phi_a(j)$ which may depend on the spin direction $a$ (this possibility was not pointed out in the mentioned references \cite{krammeretal2009,crameretal2011,haukeetal2013}); and on the conceptual side, it was inferred from our algorithm in a completely data-agnostic way, as the optimal witness for TI data at the first relaxation level of our hierarchy. In this case, the local phases are of the form $\phi_a(j)={\bf k}_a\cdot {\bf r}_j$ with ${\bf r}_j$ the position of the $j$-th subsystem. This leads to the structure factor $S_{\bf k}^{a} = N^{-1}\sum_{j,j'} e^{i{\bf k} \cdot ({\bf r}_{j'} - {\bf r}_j)}C_{jj'}^{aa}$ (notice that we have included the $j=j'$ term in the summation, corresponding to a term $C_{jj}^{aa}=1$). Although we discovered these witnesses focusing on two-dimensional systems, they can be naturally extended to arbitrary geometries. In terms of components of the structure factors, the entanglement witness of Eq.~\eqref{eq_TI_witness} reads:
\begin{equation}
	S_{{\bf k}_X}^{X} + S_{{\bf k}_Y}^{Y} + S_{{\bf k}_Z}^{Z} \ge 2 ~.
	\label{eq_TI_witness_Sk}
\end{equation}
Notice that we have defined the structure factors in terms of qubit observables, which are twice the spin observables typically used in condensed-matter physics; this leads to a factor $4$ in the definition of the structure factors. Crucially, the wavevector ${\bf k}$ may be different for the $X$, $Y$ and $Z$ components of the spins. Obviously, in order to detect entanglement, it is optimal to choose, for each spin component $a$, the direction ${\bf k}_a$ where the structure factor is \textit{minimal}, leading to:
\begin{equation}
	W_{\rm opt} = \min_{{\bf k}} S_{{\bf k}}^{X} + \min_{{\bf k}} S_{{\bf k}}^{Y} + \min_{{\bf k}} S_{{\bf k}}^{Z} \ge 2 ~.
	\label{eq_TI_witness_Sk_opt}
\end{equation}
It is in this form that the witnesses have been discovered via our algorithm, investigating the correlations generated by the dynamics in the power-law $XX$ model. We have then generalized this result to Eq.~\eqref{eq_TI_witness}. The witness Eq.~\eqref{eq_TI_witness_Sk_opt} is then violated if the fluctuations at these optimal wavevectors are suppressed below the separable bound $2$. Physically, these witnesses detect a generalized form of spin squeezing, especially suited to the many-body systems where different components of the structure factor can be measured. In several quantum simulators, the structure factors are reconstructed by Fourier transform of the real-space correlations. In condensed-matter systems, correlations are typically measured directly in momentum space via neutron scattering. During the out-of-equilibrium dynamics, these optimal wavevectors may vary over time in different ways for different spin components; and therefore the witnesses of Eq.~\eqref{eq_TI_witness}-\eqref{eq_TI_witness_Sk}-\eqref{eq_TI_witness_Sk_opt} offer a large flexibility to detect entanglement, independently of the specific SDP algorithm we used to discover them.
Finally, we notice that the witness of Eq.~\eqref{eq_TI_witness_Sk_opt} can be extended to spin-$s$ systems. We define in general the structure factor as $S_{\bf k}^{a} = N^{-1}\sum_{j,j'} e^{i{\bf k} \cdot ({\bf r}_{j'} - {\bf r}_j)}{\rm Tr}[\hat \rho \hat S_j^a \hat S_{j'}^a]$ with $\hat S_j^a$ the spin-$s$ observable in direction $a$ for subsystem $j$. Using that $Ns^2 \ge \sum_{i=1}^N \sum_{a\in\{X,Y,Z\}}|{\rm Tr}[\hat \rho \hat S_j^a]|^2$, one easily shows that $S_{{\bf k}_X}^{X} + S_{{\bf k}_Y}^{Y} + S_{{\bf k}_Z}^{Z} \ge Ns$ for all separable states. This generalizes a result of Ref.~\cite{vitaglianoetal2011} to arbitrary components of the structure factors.\\

\begin{figure}
\includegraphics[width=\linewidth]{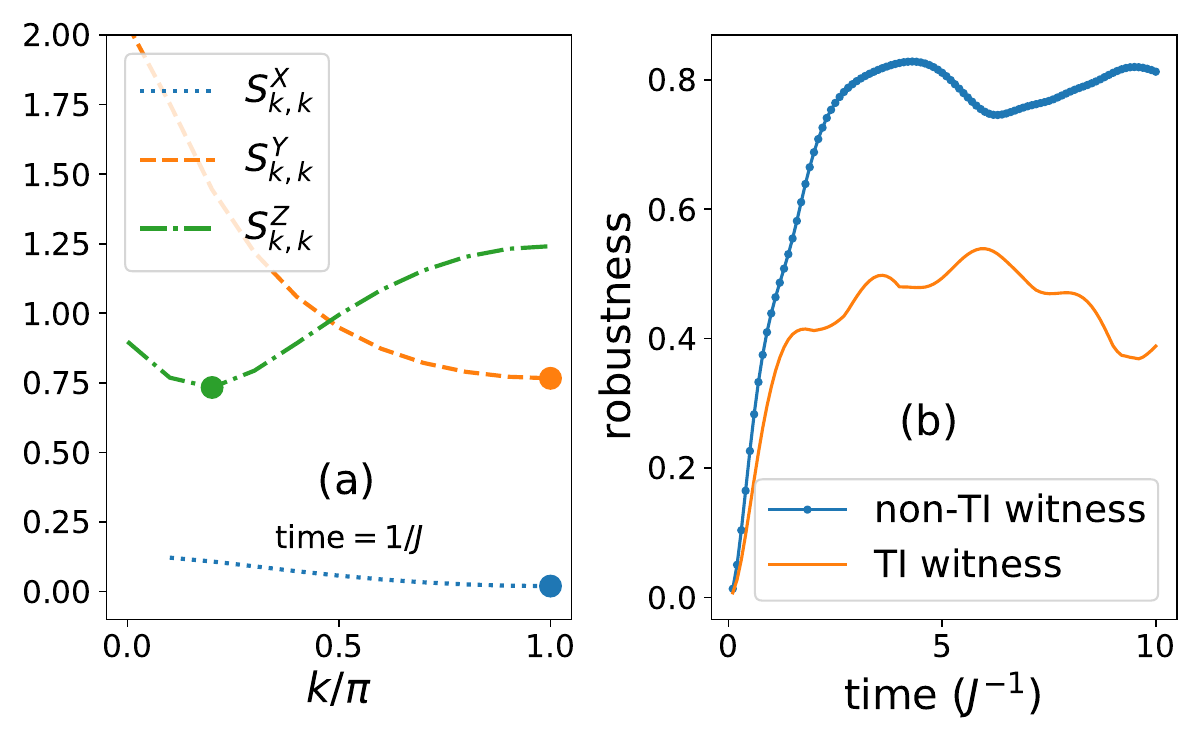}
\caption{Quench in the two-dimensional $XX$ model on a $20\times 20$ square lattice with open boundary conditions [Eq.\eqref{eq_H_XXLR_2d}]. (a) Structure factor at time $t=1/J$. The circles mark the wavevectors at which the structure factor for the $X$, $Y$ and $Z$ components is minimal. This leads to the TI-witness of Eq.~\eqref{eq_TI_witness_Sk_opt}. (b) Blue dots: noise robustness of the optimal witness found by our SDP algorithm using as input the magnetization $C_i^X$ and two-body correlations $C_{ij}^{XX}$, $C_{ij}^{YY}$ and $C_{ij}^{ZZ}$, with an initial state polarized along $x$. Orange line: noise robustness of the best translationnally-invariant (TI) witness of Eq.~\eqref{eq_TI_witness_Sk_opt}, whose reconstruction is illustrated on panel (a) for time $t=1/J$.
}
\label{fig_quench_XXLR}
\end{figure}

\noindent\textbf{Numerical results.}
In Fig.~\ref{fig_quench_XXLR}(a), we plot the structure factor at time $tJ=1$, for wavevectors ${\bf k}=(k,k)$, in a  $20 \times 20$ square lattice with open boundary conditions. The minimal value of the structure factor for each spin component (marked by a circle on the figure) then enters the TI witness of Eq.~\eqref{eq_TI_witness_Sk_opt}. In Fig.~\ref{fig_quench_XXLR}(b), we plot the noise robustness of the (non-TI) witness found by our algorithm as a function of time. As we consider open boundary conditions, the correlations have no translation invariance, and as a consequence the resulting witness loses this symmetry too. For comparison, we have also plotted the noise robustness of the TI witness of Eq.~\eqref{eq_TI_witness_Sk_opt}, evaluated at the (time-dependent) optimal wave-vectors $({\bf k}_X, {\bf k}_Y, {\bf k}_Z)$. While the TI witness reaches a noise robustness of about $0.5$, the (non-TI) optimal witness tailored to the (non-TI) correlations reaches more than $0.8$ noise robustness. However, we could not find an analytical expression for these non-TI data-driven witnesses. By a variational search over separable states, we could however verify that the separable bound obtained by our SDP algorithm was always tight.

\subsection{Bipartite witnesses}
\label{sec_bipartite_witnesses}
Finally, we show that the very same SDP algorithm outlined in Sec \ref{sec_framework} can be adapted with no additional computation cost to detect bipartite entanglement along any splitting of the system in two parts. We therefore consider a partition of the $N$ qubits into two halves $A$ and $B$, and as input data, we consider single-qubit terms $C_i^a$ and only inter-$AB$ correlations $C_{ij}^{ab}$ where $i\in A$ and $j \in B$. It is straightforward to notice that if non-full-separability can be proved from this knowledge, then the state must be bipartite entangled. Indeed, if the $AB$ state is bipartite separable: $\hat \rho^{\rm bisep}_{AB} = \sum_k p_k \hat \rho_A^{(k)} \otimes \hat \rho_B^{(k)}$, then we may define $\hat \rho^{\rm fullsep}_{AB} = \sum_k p_k \otimes_{i=1}^N \hat \rho_i^{(k)}$ with $\hat \rho_i^{(k)} = {\rm Tr}_{j \neq i}[\hat\rho_A^{(k)} \otimes \hat \rho_B^{(k)}]$. One can verify that if $\hat \rho^{\rm bisep}_{AB}$ reproduces the data, so does $\hat \rho^{\rm fullsep}_{AB}$; conversely, proving non-full-separability from these data implies bipartite entanglement. \\

\noindent\textbf{Heisenberg model and transverse-field Ising models.} We use the above idea to investigate bipartite entanglement in the transverse-field Ising and Heisenberg chain at finite temperature, for $N=64$ spins. The Heisenbeg chain is described by the Hamiltonian:
\begin{equation}
	\hat {\cal H}_{\rm Heis.} = (J/4)\sum_{i=1}^N [\hat X_i \hat X_{i+1} + \hat Y_i \hat Y_{i+1} + \hat Z_i \hat Z_{i+1}] ~,
	\label{eq_H_Heisenberg}
\end{equation}
and transverse-field Ising chain by: 
\begin{equation}
	\hat {\cal H}_{\rm Ising} = -(J/4)\sum_{i=1}^N [\hat Z_i \hat Z_{i+1} + g\hat X_i] ~.
	\label{eq_H_Ising}
\end{equation}
In Eqs.~\eqref{eq_H_Heisenberg} and \eqref{eq_H_Ising}, we have assumed periodic boundary conditions. The parameter $J$ is an overall energy scale (set to $J=1$ in our computations), and $g$ in Eq.~\eqref{eq_H_Ising} is the transverse-field amplitude. We consider thermal states $\hat \rho = Z^{-1}\exp[-\hat {\cal H} / T]$ with $T$ the temperature and $Z={\rm Tr}(\exp[-\hat {\cal H} / T])$ the partition function. As input to our algorithm, we have used all one- and two-body correlations, which are invariant under translations. These data were computed with quantum Monte Carlo. The Heisenberg model being $SU(2)$ invariant, one-body terms vanish, and two body-terms are of the form $C_{ij}^{ab} = C_{|i-j|} \delta_{ab}$ (with $a,b \in \{X,Y,Z\}$). For the Ising model the symmetries imply $C_i^Y=C_i^Z=0$, and $C_{ij}^{ab} = C^a_{|i-j|} \delta_{ab}$ (namely, off-diagonal correlations $a\neq b$ vanish). \\

\noindent\textbf{A family of bipartite entanglement witnesses.}
Having first considered a partition of the form $AAA\dots AA|BBB\dots BB$ (namely, $A=\{0,1,2, \dots, N/2-1\}$ and $B=\{N/2, \dots, N-1\}$), we have noticed that entanglement was detected iff the nearest-neighbour two-body reduced density matrix $\rho_{N/2-1, N/2}$ was itself entangled (as detected by the concurrence \cite{wooters1998}). While illustrating the relatively short-range nature of entanglement in these thermal states, we could not go beyond the mere witnessing of entanglement among nearest-neighbours. We have therefore considered a partitionning maximizing the $AB$ interface, that is: $A|B|A|B|A|B|\dots$ (namely, $A=\{0, 2, 4, \dots, N-2\}$ and $B=\{1, 3, 5, \dots, N-1\}$). This lead us to discover new  bipartite entanglement witnesses. Similarly to the case of TI multipartite entanglement witnesses [see Eq.~\eqref{eq_TI_witness}], we could analytically characterize them, and extend them to a full family of witnesses. We define:
\begin{equation}
	W_a = \sum_{i\in A} \sum_{j\in B} K_{j-i} C_{ij}^{aa} \cos[\phi_a(i) - \phi_a(j)] ~,
\end{equation}
where $\phi_a(j)$ are arbitrary local phases, and with coefficients given by:
\begin{eqnarray}
	K_r & = & \frac{2}{N} \sum_{k=-\frac{N}{4}+1}^{\frac{N}{4}-1} \exp\left(\frac{2i\pi}{N}kr\right) \\
	 &=& \frac{2}{N} \left[ \frac{\sin(\pi r / 2)}{\tan(\pi r / N)} - \cos\left(\frac{\pi r}{2}\right)\right] \\
	&\stackrel{N \gg r}{\sim}& \frac{2}{\pi r} \sin\left(\frac{\pi r}{2}\right) ~.
\end{eqnarray}
Notice that for $i\in A$ and $j \in B$, $r=j-i$ is an odd integer, in which case we have the simplified expression $K_r = \frac{2(-1)^{(r-1)/2}}{N\tan(\pi r/N)}$. The violated witnesses are then of the form
\begin{eqnarray}
	W = W_X + W_Y + W_Z \ge - \frac{N}{2} ~. \label{eq_witness_bipartite}
\end{eqnarray}
The proof of the witness inequality \eqref{eq_witness_bipartite} is given in Appendix \ref{app_bipartite_witnesses}. For the Heisenberg chain, the optimal choice of the phases is $\phi_a(i)=0$. For the Ising chain, it is $\phi_a(i)=0$ for $i \in A$, and $\phi_X(j)=\phi_Z(j)=\pi$ for $j \in B$, and $\phi_Y(j)=0$. 

\begin{figure}
\includegraphics[width=\linewidth]{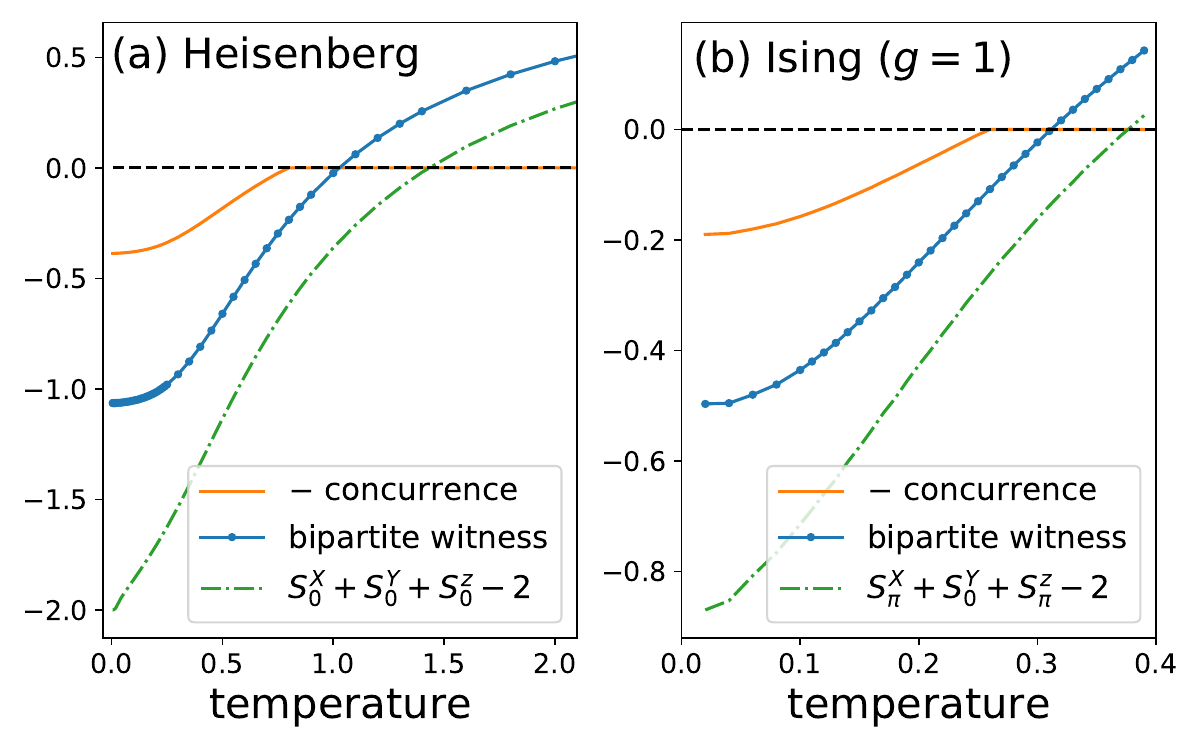}
\caption{Entanglement in (a) the Heisenberg chain [Eq.~\eqref{eq_H_Heisenberg}] and (b) the quantum Ising chain at $g=1$ [Eq.~\eqref{eq_H_Ising}], for $N=64$ spins. The bipartition is of the form $A|B|A|B|A|B\dots$. Solid orange line: (minus) the concurrence \cite{wooters1998} between nearest-neighbours. Blue dots: bipartite witness $1+(2W/N)$ with $W$ given in Eq.~\eqref{eq_witness_bipartite}. Dashed-dotted green line: multipartite witness $S_{k_X}^X + S_{k_Y}^Y + S_{k_Z}^Z - 2$ at the optimal wavevectors $k_a$ [see text and Eq.~\eqref{eq_TI_witness_Sk_opt}] (for all witnesses, entanglement is detected for negative values). 
}
\label{fig_bipartite_Heisenberg_Ising}
\end{figure}

As illustrated in Fig.~\ref{fig_bipartite_Heisenberg_Ising}, the bipartite entanglement witnesses of Eq.~\eqref{eq_witness_bipartite} allow one to detect entanglement in regimes where all two-body reduced density matrices are separable (as measured by the concurrence \cite{wooters1998}).\\

\noindent\textbf{Comparison with the structure factor witnesses.} For the sake of completeness, we have also evaluated the multipartite entanglement witness based on the structure factor [Eq.~\eqref{eq_TI_witness_Sk_opt}] discussed in Section \ref{sec_2dXX}. As this witness involves also intra-$A$ and intra-$B$ correlations, and is based on the same SDP criterion, it detects entanglement at strictly higher temperatures than the bipartite witness of Eq.~\eqref{eq_witness_bipartite}.

The optimal wavevectors $k_X$, $k_Y$ and $k_Z$ are found from the following observations. The Heisenberg model develops antiferromagnetic correlations at low temperature, leading to a peak at $k=\pi$ in the structure factor. Concomitantly, fluctuations of the uniform magnetization (at $k=0$) are suppressed, reaching a spin-singlet state in the ground state ($S_0^a=0$). 
At all temperatures, the structure factor is always minimal at $k_X = k_Y = k_Z = 0$, and in this case the optimal witness is simply $\sum_a S_0^a \ge 2$. As previously reported in Ref.~\cite{frerotR2021}, it is violated up to a temperature $T/J \approx 1.4$ [Fig.~\ref{fig_bipartite_Heisenberg_Ising}(a)]. In contrast, at low temperature and around the quantum critical point $g=1$ \cite{sachdev2011quantum}, the Ising model develops ferromagnetic correlations for the $Z$ component of the spin, and the structure factor is minimal at $k_Y=\pi$. The uniform magnetization along $Y$ is slightly squeezed below the standard quantum limit \cite{frerotR2018_1}, and the structure factor is minimal at $k_Y=0$. Finally, correlations in the $X$ direction (the direction of the transverse-field) are strongly ferromagnetic, and are suppressed at $k_X=\pi$. We have found that these choices are optimal throughout the phase diagram. As illustrated in Fig.~\ref{fig_bipartite_Heisenberg_Ising}(b), above the quantum critical point $g=1$ the optimal witness $S_\pi^X+S_0^Y+S_\pi^Z\ge 2$ is violated for temperatures $T/J \lesssim 0.38$. In contrast a criterion based on the quantum Fisher information (based on the dynamical structure factor for $ZZ$ correlations \cite{haukeetal2016}, and which is considerably more challenging to estimate, both in theory and in experiments \cite{mathewetal2020,scheieetal2021}) is violated for $T/J \lesssim 0.11$; and a witness based on the same data as used in the present work \cite{frerotR2021}, and optimized for $T/J=0.28$, is violated for $T/J\lesssim 0.31$. The entanglememt witness of Eq.~\eqref{eq_TI_witness_Sk_opt} discovered in the present work offers therefore both a simpler and more robust criterion for the detection of multipartite entanglement in many-body systems, as compared to existing criteria proposed so far in the literature.

\section{Conclusion}
\label{sec_conclusion}
We have implemented a systematic, scalable and flexible approach to detect multipartite entanglement in many-body systems. Assuming that the knowledge of some average values of many-body observables are known, one can build a correlation matrix whose entries reproduce these data. Under the assumption that the state is separable, positive semidefinite constraints must be obeyed by the correlation matrix. Verifying these constraints is efficiently achieved via semidefinite programming techniques, yielding a data-tailored entanglement witness violated by the data under consideration. We have illustrated the scalability of this approach in some paradigmatic examples of many-body systems, demonstrating for instance how our approach can easily deal with systems of hundreds of qubits when two-body correlations are used as input data.  

By choosing to perform the proposed entanglement test as a noise robustness problem, we were able to show that the corresponding entanglement witnesses can tolerate realistic amount of noise in many physically relevant many-body scenarios. We have also shown that the specific entanglement witnesses discovered via our approach can sometimes be analyzed analytically, leading to explicit entanglement witness of independent relevance [see e.g. Eq.~\eqref{eq_TI_witness}, \eqref{eq_TI_witness_Sk_opt} or \eqref{eq_witness_bipartite}]. 

Within our framework, one can probe the (in)compatibility of a given set of average values with a separable state for any fixed partitioning of the system. Here, we focused mostly on a partitioning into $N$ individual qubits, or a bipartition into two halves of $N/2$ qubits. {Our approach can be used to detect entanglement among individually addressed qubits, as extensively demonstrated throughout the paper. It can also be used to detect entanglement among indistinguishable particles, probed e.g. via collective spin observables; as a matter of fact, all generalized spin-squeezing inequalities \cite{tothetal2009} typically used to detect entanglement in this framework are recovered as a special case by our method.} It is \textit{a priori} unclear if our approach can be extended to encompass also statistical mixtures of different partitionings, as required to quantify the so-called entanglement depth, or width \cite{guhneT2009}. 

Several research directions are now open to future works. Fist,  the considered approach can be extended to qudit systems in a straightforward manner, which provides a scalable technique for the investigation of multipartite entanglement in large ensembles of qudits. Then, although we have illustrated the method by assuming that one- and two-body correlations are known, one could naturally include the knowledge of any $k$-point function. { Including such higher-order correlations would certainly enhance the capability to detect entanglement in the examples we have presented.} Studying topological phases \cite{wen_choreographed_2019}, where entanglement could be revealed by string-order-parameters, via specific entanglement witnesses inferred by our data-driven method, represents also an exciting avenue for future works. Finally, implementing our algorithm using experimental data as input would probably reveal unforeseen features of many-body entanglement. This last possibility is especially relevant in the context of quantum computation and quantum simulation, operating beyond the capabilities of classical computers \cite{Martinis2019,chiuetal2019,koepselletal2019,ebadi2020quantum,zhong_quantum_2020,scholl2020programmable,Bluvstein_2021}, and where entanglement is commonly assumed to be an essential resource \cite{Preskill2018quantumcomputingin}.  

\acknowledgments{We thank Tommaso Roscilde for providing us the quantum Monte Carlo data used in Section IV.C. We thank Felix Huber for drawing our attention to ref.~\cite{bohnet_waldraff_etal_2017} after the completion of the first version of this paper. This work is supported by the ERC AdG CERQUTE, the AXA Chair in Quantum Information Science, the Government of Spain (FIS2020-TRANQI, Severo Ochoa CEX2019-000910-S and Retos QuSpin), Fundacio Cellex, Fundacio Mir-Puig, Generalitat de Catalunya (CERCA, AGAUR SGR 1381 and QuantumCAT), the Austrian Science Fund (FWF) through Project number 414325145 within SFB F7104 and the Alexander von Humboldt Foundation, the Agence Nationale de la Recherche (ANR) Research Collaborative Project Qu-DICE (ANR-PRC-CES47), the John Templeton Foundation (Grant No. 61835).}

\bibliography{biblio}
\appendix

\section{Technical considerations}
\label{sec_technical}

\subsection{Symmetrizing the unknown $p$ distribution}
\label{app_symmetries}
In this section, we explain how the number of free variables in the SDP algorithm solving Eq.~\eqref{eq_Gamma_matrix} can be drastically reduced by symmetrizing the underlying $p(\{x_i, y_i, z_i\})$ distribution. This symmetrization only relies on the nature of the data themselves, and does not assume that the quantum state has specific symmetries.

1) First, let us assume that the available data consist of $\{C_i^Z, C_{ij}^{XX}, C_{ij}^{YY}, C_{ij}^{ZZ}\}$, namely the magnetization along the direction $Z$ for all qubits, and pair correlations along the same direction. The data we used in analyzing the quench in the two-dimensional $XX$ model with $1/r^3$ interactions, and thermal states of the quantum Ising model, have this structure (exchanging the role of the $Z$ and $X$ direction). Let us assume that a distribution $p_0$ exists which reproduces the data. One may then consider the distributions $p_1 = p_0(\{-x_i, y_i, z_i\})$, $p_2 = p_0(\{x_i, -y_i, z_i\})$, and $p_3 = p_0(\{-x_i, -y_i, z_i\})$. Clearly, one has that $\langle z_i \rangle_{p_0} = \langle z_i \rangle_{p_k}$ for $k \in \{1,2,3\}$, and similarly for two-body terms: $\langle a_i a_j \rangle_{p_0} = \langle a_i a_j \rangle_{p_k}$ for $a \in \{x,y,z\}$ and $k \in \{1, 2, 3\}$. In other words, $p_0$ reproduces the data iff $p_k$ reproduces the data for $k \in \{1,2,3\}$. One may then consider the distribution $p_4 = (p_0 + p_1 + p_2 + p_3) / 4$, which also reproduces the data. Importantly, $p_4$ is such that $\langle x_i \rangle_{p_4} = 0$ (since under $p_4$, $x_i$ has the same probability as $-x_i$). Similarly,  $\langle y_i \rangle_{p_4} = 0$, and two-body terms are such that $\langle x_i y_j \rangle_{p_4} = \langle x_i y_j \rangle_{p_4} = \langle y_i z_j \rangle_{p_4} = 0$.
 Therefore, without loss of generality, one may impose in the SDP algorithm that $\langle x_i \rangle = \langle y_i \rangle = \langle x_i y_j \rangle=\langle x_i z_j \rangle= \langle y_i z_j\rangle = 0$ for all $i,j$. One thus reduces the number of free variables in Eq.~\eqref{eq_Gamma_matrix} from order $O(N^2)$ to $2N$ (for instance, all $\langle x_i^2 \rangle$ and $\langle y_i^2 \rangle$ for $i \in [N]$).\\
 
2) Let us then consider a situation where the data consist of $\{C_i^Z, C_{ij}^{XX} + C_{ij}^{YY}, C_{ij}^{ZZ}\}$, as was the case in the study of the quench in the one-dimensional $XX$ chain. Let us assume that a probability distribution $p(\{x_i, y_i, z_i\})$ reproduces the data. We consider then the distribution $p'=p(\{y_i, x_i, z_i\})$. As $\langle x_i x_j + y_i y_j \rangle_{p} = \langle x_i x_j + y_i y_j\rangle_{p'}$, this distribution also reproduces the data. Hence, we may as well consider $q=(p+p')/2$, which also reproduces the data. As $q$ is such that $\langle x_i x_j \rangle_q = \langle y_i y_j \rangle_q$ for all pairs $(i,j)$, we may assume $C_{ij}^{XX}=C_{ij}^{YY}$ in the SDP algorithm without loss of generality, and without assuming that this symmetry is actually present in the experiment. Following similar arguments, the distribution $p''=p(\{-x_i, y_i, z_i\})$ reproduces the data, since $\langle x_i x_j \rangle_{p} = \langle x_i x_j \rangle_{p''}$. Considering $(p+p'')/2$, we may impose $\langle x_i \rangle = 0$, and also $\langle x_i y_j \rangle=\langle x_i z_j \rangle=0$ for all pairs $(i,j)$. The same reasoning apply interchanging the role of $x$ and $y$, leading to $\langle y_i\rangle=\langle y_i z_j \rangle = 0$. In conclusion, we may solve the SDP in the form of Eq.~\eqref{eq_Gamma_matrix} imposing that all terms are zero, except the data $C_i^Z$, $C_{ij}^{XX}=C_{ij}^{YY}$ [obtained as $(C_{ij}^{XX}+C_{ij}^{YY}) / 2$], and $C_{ij}^{ZZ}$; and except the diagonal terms with $\langle x_i^2 \rangle = \langle y_i^2\rangle$. This reduces the number of free variables from order $O(N^2)$ to $N$ (for instance, all $\langle z_i^2 \rangle$ for $i \in [N]$), greatly improving the scalability. \\

3) Finally, if one only knows $c_{ij} := C_{ij}^{XX} + C_{ij}^{YY} + C_{ij}^{ZZ}$, by a similar reasoning one may consider a distribution $p(\{x_i, y_i, z_i \})$ which is invariant under all rotations, satisfying $\langle x_i \rangle = \langle y_i \rangle = \langle z_i \rangle = 0$, $\langle x_i y_j \rangle = \langle x_i z_j \rangle = \langle y_i z_j \rangle = 0$, and $\langle x_i x_j \rangle = \langle y_i y_i \rangle = \langle z_i z_j \rangle$. In this case, there is no free variable at all in Eq.~\eqref{eq_Gamma_matrix} [all two-body diagonal terms are $1/3$, and the only non-zero off-diagonal entries are $C_{ij}^{XX}=C_{ij} ^{YY} = C_{ij}^{ZZ}$, obtained as $c_{ij} / 3$]. As was illustrated in the case of the Werner state, testing entanglement via Eq.~\eqref{eq_Gamma_matrix} simply consists of checking positive semidefiniteness of the matrix $M$, defined by $M_{ij} = c_{ij}$ if $i \neq j$, and $M_{ii}=1$. We emphasize that this does not assume any spatial symmetry, and applies in particular to data with no translation invariance.

\subsection{Robust entanglement witness from one- and two-body correlations on $N$ qubits}
\label{sec_witness_qubits}
Here we provide details on the way to extract an entanglement witness of the form \eqref{eq_witness} via semidefinite programming. Specifically, we show how to find the \textit{noise robustness} of the entanglement contained in the data, and derive an entanglement witness via semidefinite programming. We start considering a situation where $N$ qubits are measured, and some of the one-body terms $C_i^a$ and two-body terms $C_{ij}^{ab}$ have been measured. We denote generically these data as $\{C_\alpha\}_{\alpha=1}^R:=\{C_i^a, C_{ij}^{ab}\}$, where $\alpha$ labels both the sites and measurements. We assume that in total the dataset contains $R$ single- and two-body correlations. We define the noise robustness as the minimal value of $\lambda$ such that $\{(1-\lambda)C_\alpha\}$ is compatible with a positive semidefinite (PSD) correlation matrix, as explained in the main text [Section \ref{sec_framework} and in particular Eq.~\eqref{eq_Gamma_matrix}]. That is, we aim at solving the problem:
\begin{subequations}
\label{eq_SDP_qubits_1}
\begin{align}
	\min_{\lambda \ge 0} &\quad \lambda \quad {\rm s.t.}  \nonumber \\
	\text{(PSD)}&\quad 	\Gamma \succeq 0 \label{eq_SDP_qubits_1_gamma}\\
	\text{(data)}&\quad 	\Gamma_{(n,m)(\alpha)} = (1-\lambda)C_\alpha &(\alpha \in [R]) \label{eq_SDP_qubits_1_data}\\
	\text{(Pauli)} &\quad 	\sum_{a=1}^3 \Gamma_{3(i-1)+a,3(i-1)+a} = 1 & (i \in [N]) \label{eq_SDP_qubits_1_Pauli}
\end{align}
\end{subequations}
The (symmetric) matrix $\Gamma=(\Gamma_{n,m})_{0\le n,m\le 3N}$ is the correlation matrix for the variables $\{1\}\cup\{x_i, y_i, z_i\}_{i=1}^N$ parametrizing separable states of $N$ qubits (Section \ref{sec_framework}), and is therefore PSD [Eq.~\eqref{eq_SDP_qubits_1_gamma}]. In Eq.~\eqref{eq_SDP_qubits_1_data}, we constrain the relevant entries of the $\Gamma$ matrix to reproduce the data (with a $1-\lambda$ noise prefactor); we have introduced $(n,m)(\alpha)$ to denote the pair of indices $(n,m)$ containing the data $C_\alpha$. Specifically [see Eq.~\eqref{eq_Gamma_matrix}], $C_i^a$ is contained in $\Gamma_{0,3(i-1)+a}$, and $C_{ij}^{ab}$ in $\Gamma_{3(i-1)+a, 3(j-1)+b}$. Finally, Eq.~\eqref{eq_SDP_qubits_1_Pauli} enforces $x_i^2+y_i^2+z_i^2=1$ in the $\Gamma$ matrix; as this condition descends from properties of the Pauli matrices (Section \ref{sec_framework}), we call it the Pauli constraint.\\

\noindent\textit{Standard primal form.}
We now rewrite in its so-called standard primal form \cite{parillo_book} the semidefinite program (SDP) defined in Eq.~\eqref{eq_SDP_qubits_1}:
\begin{align}
\min_{X \succeq 0} & \quad \langle M , X \rangle  \quad \text{s.t.} \nonumber \\
\text{(data)}&  \quad \langle A_\alpha^{\rm data} , X \rangle = C_\alpha & (\alpha \in [R]) \label{eq_SDP_qubits_primal}\\
\text{(Pauli)}&   \quad \langle A_i^{\rm Pauli} , X \rangle = 1 & (i \in [N]) \nonumber
\end{align}
We have introduced the matrix scalar product $\langle X,Y\rangle = {\rm Tr}(X^T Y) = \sum_{ij}X_{ij}Y_{ij}$. The involved matrices have the following block-diagonal form: 
\begin{subequations}
\label{eq_def_MXA_matrices}
\begin{align}
&M = \left( 
		\begin{array}{c|c}
			1 &  0 \\
			\hline
			0 & 0
		\end{array}
	\right),~
X = \left( 
		\begin{array}{c|c}
			\lambda &  0 \\
			\hline
			0 & \Gamma
		\end{array}
	\right),~\\
&A_\alpha^{\rm data} = \left( 
		\begin{array}{c|c}
			C_\alpha &  0 \\
			\hline
			0 & \Sigma_\alpha^{\rm data}
		\end{array}
	\right),~
A_i^{\rm Pauli} = \left( 
		\begin{array}{c|c}
			0 &  0 \\
			\hline
			0 & \Sigma_i^{\rm Pauli}
		\end{array}
	\right)
\end{align}
\end{subequations}
where the $\Sigma$ matrices are given by:
\begin{subequations}
\label{eq_def_Sigma_matrices}
\begin{align}
&\left( \Sigma_\alpha^{\text{data}} \right)_{n,m} = \left\lbrace
\begin{array}{cc}
1 ~\text{if}~(n,m)=(n,m)(\alpha)  \\ 
0 ~\text{otherwise}
\end{array}
\right. \\
&\left(
	\Sigma_i^{\text{Pauli}} 
\right)_{n,m} = \sum_{a = 1}^3 \delta_{n,m} \delta_{m,3(i-1) + a } 
\end{align}
\end{subequations}
The SDP \eqref{eq_SDP_qubits_primal} is equivalent to the problem \eqref{eq_SDP_qubits_1} of finding the minimal noise $\lambda$ for which the $\Gamma$ matrix in \eqref{eq_Gamma_matrix} becomes positive semidefinite, for a set of noisy data $\lbrace (1-\lambda) C_i^{a} , (1-\lambda) C_{ij}^{ab}\rbrace$.
Therefore, an optimal solution $\lambda^\ast > 0$ implies that the given data are not compatible with a separable state, hence resulting in entanglement detection. \\

\noindent\textit{Dual form.}
If that is the case, one can derive an entanglement witness by considering the so-called dual problem \cite{parillo_book} corresponding to Eq.~\eqref{eq_SDP_qubits_primal}:
\begin{align}
\label{eq_SDP_qubits_dual_0}
&\max_{\vec{w}, \vec{w}^{\rm Pauli}} \quad \sum_{\alpha=1}^R w_{\alpha} C_\alpha + \sum_{i=1}^N w_{i}^{\text{Pauli}}\nonumber \quad \text{s.t.} \\
& \sum_{\alpha=1}^R w_{\alpha}  A_{\alpha}^{\text{data}} + \sum_{i=1}^N w_{i}^{\text{Pauli}}  A_{i}^{\text{Pauli}}\preceq M ~.
\end{align}
Using the expressions of the matrices $M$ and $A$'s [Eqs.~\eqref{eq_def_MXA_matrices} and \eqref{eq_def_Sigma_matrices}], we rewrite Eq.~\eqref{eq_SDP_qubits_dual_0} as:
\begin{align}\label{eq_SDP_qubits_dual}
&\max_{\vec{w}, \vec{w}^{\rm Pauli}} \quad \quad \vec{w} \cdot \vec{C}  + \sum_{i=1}^N w_i^{\text{Pauli}} \quad \text{s.t.} \nonumber \\
&\vec{w} \cdot \vec{C}  \leq 1 \\
& \sum_{\alpha=1}^R w_{\alpha}  \Sigma_{\alpha}^{\text{data}} + \sum_{i=1}^N w_{i}^{\text{Pauli}}  \Sigma_{i}^{\text{Pauli}}  \preceq 0 \, . \nonumber
\end{align}
We denote $\{\vec{w}^\ast, (\vec{w}^{\rm Pauli})^\ast\}$ the optimal solution to this dual problem.\\

\noindent\textit{Strong duality.}
As for all semidefinite progams \cite{parillo_book}, any primal feasible $X$ [that is, a PSD matrix $X \succeq 0$ satisfying the constraints in Eq.~\eqref{eq_SDP_qubits_primal}] yields an upper bound to any dual feasible [that is, any $\{w_\alpha, w_i^{\rm Pauli}\}$ satisfying the constraints in Eq.~\eqref{eq_SDP_qubits_dual_0}]. Indeed, for any such feasible $X$ and $\vec w$, we have: $\langle M,X\rangle - \sum_{\alpha} w_\alpha C_\alpha - \sum_i w_i^{\rm Pauli} = \langle (M - \sum_\alpha w_\alpha A_\alpha^{\rm data} - \sum_i w_i^{\rm Pauli} A_i^{\rm Pauli}), X\rangle$, where we used the constraints in Eq.~\eqref{eq_SDP_qubits_primal}. Using then the constraint in Eq.~\eqref{eq_SDP_qubits_dual_0}, we notice that $(M - \sum_\alpha w_\alpha A_\alpha^{\rm data} - \sum_i w_i^{\rm Pauli} A_i^{\rm Pauli}) \succeq 0$. As $X\succeq 0$, and using the fact that $\langle Y,X \rangle \ge 0$ for any two PSD matrices $Y$ and $X$, we conclude that $\langle M,X\rangle - \sum_{\alpha} w_\alpha C_\alpha - \sum_i w_i^{\rm Pauli} \ge 0$. In particular, the primal optimum upper bounds the dual optimum: $\lambda^\ast \ge \vec{w}^\ast \cdot \vec C + \sum_i (w_i^{\rm Pauli})^\ast$, a property known as \textit{weak duality} \cite{parillo_book}. In our case, the primal and dual optima are actually equal, as a consequence of \textit{strong duality} which holds for our problem. A sufficient condition for strong duality to hold \cite{parillo_book} is that both primal and dual problems are \textit{strictly} feasible. The primal problem is stricly feasible if one can find a positive definite matrix $X \succ 0$ satisfying the constraints in Eq.~\eqref{eq_SDP_qubits_primal}: such strictly feasible $X$ is readily obtained by choosing $\lambda = 1$ and $\Gamma=\mathbb{1}/3$ in Eq.~\eqref{eq_def_MXA_matrices} [or, equivalently, in Eq.~\eqref{eq_SDP_qubits_1}]. The dual problem is stricly feasible if one can exhibit some $\{w_\alpha, w_i^{\rm Pauli}\}$ satisfying the constraints in Eq.~\eqref{eq_SDP_qubits_dual} as strict inequalities: such stricly feasible $w$'s are readily obtained as $w_\alpha=0$ and $w_i^{\rm Pauli}=-1$ (such that $\sum_i w_i^{\rm Pauli} \Sigma_{i}^{\rm Pauli}=-\mathbb{1}$). Two important properties follow from strong duality: 1) $\lambda^\ast = \vec{w}^\ast \cdot \vec C + \sum_i (w_i^{\rm Pauli})^\ast$; 2) as a consequence, we also have $\langle (M - \sum_\alpha w_\alpha^\ast  A_\alpha^{\rm data} - \sum_i (w_i^{\rm Pauli})^\ast  A_i^{\rm Pauli}), X^\ast \rangle = 0$. Using the expression of $M$ and the $A$'s matrices [Eq.~\eqref{eq_def_MXA_matrices}], this implies in particular $(1 - \vec{w}^\ast \cdot \vec C)\lambda^\ast = 0$. Whenever entanglement is detected ($\lambda^\ast > 0$), we have therefore $\vec{w}^\ast \cdot \vec C=1$, so that $-\sum_i (w_i^{\rm Pauli})^\ast = 1 - \lambda^\ast$.  \\

\noindent\textit{Entanglement witness.} The coefficients $w_\alpha^\ast$ define an entanglement witness whose separable bound is given by $1-\lambda^\ast$:
\begin{equation}
	\sum_\alpha w_\alpha^\ast C_\alpha^{\rm sep} \leq 1 - \lambda^\ast ~,
	\label{eq_witness_from dual}
\end{equation}
which holds for all separable data $\{C_\alpha^{\rm sep}\}$ (namely, data compatible with a separable state), and which is violated by the data under consideration ($\sum_\alpha w_\alpha^\ast C_\alpha = 1)$. In order to prove this fact, we first observe that $\vec{w}^\ast \cdot \vec{C}^{\rm sep} \le 1$ for all separable data $\vec{C}^{\rm sep}$ (indeed: $\vec{w}^\ast \cdot \vec{0} = 0$ for the separable data $\vec{0}$, while $\vec{w}^\ast \cdot \vec{C} = 1$ for the non-separable data $\vec{C}$; we conclude using the convexity of the set of separable data). Therefore, $\{\vec{w}^\ast, (w_i^{\rm Pauli})^\ast\}$ represent a dual feasible [Eq.~\eqref{eq_SDP_qubits_dual}] for any separable data $\vec{C}^{\rm sep}$. As a consequence of duality, this provides a lower bound to the primal optimum, which is $(\lambda^*)_{\rm sep}=0$ for separable data. Hence, $\sum_\alpha w_\alpha^\ast C_\alpha^{\rm sep} + \sum_i (w_i^{\rm Pauli})^\ast \le 0$, namely Eq.~\eqref{eq_witness_from dual}.

\subsection{General formulation of the hierarchy}
\label{sec_general_formulation}

\noindent\textit{Building the correlation matrix.} The core of our approach is to build a suitable correlation matrix $\Gamma$ for the local classical variables $\{{\bf n}_i\}$ with ${\bf n}_i=(x_i,y_i,z_i)$, akin to Eq.~\eqref{eq_Gamma_matrix}. The following conditions are necessarily fulfilled for all separable data (that is, data compatible with a separable state): 1) semidefinite positivity of $\Gamma$; 2) compatibility of $\Gamma$ with the available data $\{C_\alpha\}$; and 3) compatibility of the local variables $\{{\bf n}_i\}$ with local qubit states (i.e. $x_i^2 + y_i^2 + z_i^2 = 1$). Crucially, verifying conditions 1-3 can be cast in the form of a semidefinite program. Generically, the $\Gamma$ matrix is built as:
\begin{subequations}
	\label{eq_def_gamma_generic}
\begin{align}
	&\Gamma = \langle {\bf v}^T {\bf v} \rangle \\
	&{\bf v} = [f_1(\{{\bf n}_i\}), \dots, f_M(\{{\bf n}_i\})] ~,
\end{align}
\end{subequations}
where the $f_i$ are arbitrary polynomial functions of the local variables ${\bf n}_i$. In practice, the choice of these functions influences both the tightness and the computational cost of the approach, and considerable flexibility is offered to find the best tradeoff between them -- with the constraint that the data $\{C_\alpha\}$ can be expressed as linear combinations of the entries of the $\Gamma$ matrix. Throughout this work we considered the following simple choice dictated by the nature of the data we aimed at reproducing:
\begin{equation}
	{\bf v}^{(1)} = \{1\} \cup \{n_a^{(i)}; 1 \le i \le N; a \in \{x,y,z\}\} ~.
	\label{eq_level_1_monomials}
\end{equation}
As we discuss below, this defines the first level of a (convergent) hierarchy.
\\

\noindent\textit{Expression of the semidefinite program.} For any choice of polynomial functions $f_i$ in Eq.~\eqref{eq_def_gamma_generic}, one may parallel the procedure described in details for qubits in Section \ref{sec_witness_qubits}, leading a semidefinite program of the form:
\begin{subequations}
 \label{eq_SDP_qudits_primal}
\begin{align}
\min_{X \succeq 0} & \quad \langle M , X \rangle  \quad \text{s.t.} \label{eq_SDP_qudits_PSD_condition}\\
\text{(data)}&  \quad \langle A_\alpha^{\rm data} , X \rangle = C_\alpha & (\alpha \in [n_{\rm data}])\label{eq_SDP_qudits_data_conditions} \\
\text{(Pauli)}&   \quad \langle A_i^{\rm Pauli} , X \rangle = b_i & (i \in [n_{\rm Pauli}]) \label{eq_SDP_qudits_GM_conditions}
\end{align}
\end{subequations}
where the matrices $X$, $M$, $A$'s are as in Eq.~\eqref{eq_def_MXA_matrices}. The expression of the matrices $\Sigma_\alpha^{\rm data}$, $\Sigma_i^{\rm Pauli}$, and of the parameters $b_i$, depend on the specific choice of the monomials. Condition \eqref{eq_SDP_qudits_data_conditions} ensures that certain linear combinations of entries of the correlation matrix $\Gamma$ reproduce the (noisy) data $(1-\lambda)C_{\alpha}$ (it is possible that several independent linear combinations reproduce the same data; hence, the number $n_{\rm data}$ of such constraints may be larger than the total number $R$ of data $C_\alpha$). Conditions \eqref{eq_SDP_qudits_GM_conditions} enforce the compatibility of the local variables $\{{\bf n}_i\}$ with local qubit states.\\

\noindent\textit{Dual problem and entanglement witness.}
Exactly as for the case discussed in Section \ref{sec_witness_qubits}, if the (primal) problem \eqref{eq_SDP_qudits_primal} is unfeasible, an optimal $\lambda^*>0$ is obtained. One may then derive an entanglement witness by solving the (dual) problem:
\begin{align}\label{eq_SDP_qudits_dual}
&\max_{\vec{w}, \vec{w}^{\rm Pauli}} \quad \quad \vec{w} \cdot \vec{C}  + \sum_i b_i w_i^{\text{Pauli}} \quad \text{s.t.} \nonumber \\
&\vec{w} \cdot \vec{C}  \leq 1 \\
& \sum_{\alpha} w_{\alpha}  \Sigma_{\alpha}^{\text{data}} + \sum_{i} w_{i}^{\text{Pauli}}  \Sigma_{i}^{\text{Pauli}}  \preceq 0 \, . \nonumber
\end{align}
If strong duality holds (see Section \ref{sec_witness_qubits}), the optimal coefficients $\vec{w}^*$ allow one to build an entanglement witness as in Eq.~\eqref{eq_witness_from dual}. The violation of this witness by the data under consideration ultimately certifies the presence of entanglement in the system.\\

\noindent\textit{Convergence of the hierarchy.} Each choice of polynomial functions $f_i$ in Eq.~\eqref{eq_def_gamma_generic} defines a different relaxation to the set of separable data. One may actually formulate a systematic hierarchy of such choices, converging towards the exact separable set. The $l$-th relaxation level is defined as:
\begin{eqnarray}
	&{\bf v}^{(0)} = \{1\} \nonumber \\
	&{\bf u}^{(l)} = \left\{
	\prod_{r=1}^l n_{a_r}^{(i_r)};
		~a_r \in \{x,y,z\}; 
		~ 1 \le i_1 \le \dots \le i_l \le N
	\right\} \nonumber \\
	& {\bf v}^{(l)} = {\bf v}^{(l-1)} \cup {\bf u}^{(l)} ~.\label{eq_def_relaxation_monomials}
\end{eqnarray}
Throughout this work, we considered only the first relaxation level defined by Eq.~\eqref{eq_level_1_monomials}. Crucially, the corresponding hierarchy converges, in the limit $l \to \infty$, towards the separable set. A way to see that is to interpret the hierarchy as Lassere's series of relaxation for the moment problem associated to the variables $\lbrace(x_i,y_i,z_i)\rbrace$. Since the variables satisfy the quadratic constraint \eqref{eq_local_constraint_LV}, the relaxation meets the Archimedean condition, which is enough to guarantee convergence of the hierarchy \cite{lasserre2001global}. It follows that the set of correlations that can be recovered as moments of an overall distribution $p(\{{\bf n}_i\})$, i.e. the set of separable correlations, is obtained as the asymptotic limit of the hierarchy defined above. Therefore, if the data are incompatible with a separable state, they will be detected as entangled at a finite level of the hierarchy -- althought the computation cost of high level tests quickly increases with $l$, as one needs to manipulate a correlation matrix of size $\sim (4N)^l$ in the semidefinite program.

Lastly, notice that one can straightforwardly define some hybrid levels of the hierarchy, where the entries of the vector ${\bf v}^{(l)}$ are complemented by monomials of order higher than $l$. Such hybrid conditions have the flexibility of including the knowledge of a finite amount of higher-order correlations, while retaining the scalability of the computational cost given by the fixed level $l$. \\

\noindent\textit{Invariance of the hierarchy under partial transposition.} In Section \ref{sec_framework} of the main text, we already noticed that the level-1 relaxation is left invariant by the partial transposition (PT) of any subsystem. The key observation was that partial transposition simply amounts to a change of basis for the $\Gamma$ matrix -- hence, positivity of $\Gamma$ is left unchanged under PT. The same observation carries over to arbitrary relaxation levels [Eq.~\eqref{eq_def_relaxation_monomials}]. Indeed, the Pauli matrices are either symmetric or antisymmetric; therefore, under PT they are either left invariant, or transformed into their opposite. In terms of the correlations of the ${\bf n}_i$ variables [cf.~Eq.~\eqref{eq_Gamma_matrix}], or in terms of the choice of monomials at a given relaxation level [cf.~Eq.~\eqref{eq_def_relaxation_monomials}], it simply amounts to change the corresponding variables into their opposite, which is again achieved by a change of basis. Our approach therefore represents a hierachy of conditions which are completely independent of the PPT-based criteria \cite{Peres1996,dohertyetal2005,neven2021symmetryresolved,yu2021optimal}.

\section{Proof of the bipartite entanglement witness}
\label{app_bipartite_witnesses}

Here we prove the validity of \eqref{eq_witness_bipartite} as a witness of bipartite entanglement according to an even-odd partitioning of the system. For the sake of completeness, we first revise the setting and the witness expression.
We consider the bipartition $A=\{0, 2, 4, \dots, N-2\}$ and $B=\{1, 3, 5, \dots, N-1\}$. We introduce local phases $\phi_a(i)$ for $a \in \{X, Y, Z\}$ and $i \in [N]$, and define:
\begin{equation}
	2W_a = \sum_{j\in A} \sum_{j'\in B} K_{j-j'} C_{jj'}^{aa} e^{i[\phi_a(j) - \phi_a(j')]} + {\rm c.c.}~.
\end{equation}
The coefficients $K_r$ are given by:
\begin{eqnarray}
	K_r = K_{-r} & = & \frac{2}{N} \sum_{k=-\frac{N}{4}+1}^{\frac{N}{4}-1} \exp\left(\frac{2i\pi}{N}kr\right) \\
	 &=& \frac{2}{N} \left[ \frac{\sin(\pi r / 2)}{\tan(\pi r / N)} - \cos\left(\frac{\pi r}{2}\right)\right] ~. \label{eq_expression_Kr_bipartite}
\end{eqnarray}
The violated witnesses are then of the form:
\begin{equation}
	W_X + W_Y + W_Z \ge -\frac{N}{2} ~. \label{eq_witness_bipartite_appendix}
\end{equation}
\begin{proof}
The proof of this inequality is as follows. Assuming that the state is fully separable, we have:
\begin{eqnarray}
	2W_a & = &\sum_{j,j'=0}^{N-1} K_{j-j'} \langle a_j a_{j'}  \rangle e^{i[\phi_a(j) - \phi_a(j')]} - \nonumber \\ 
	&& \sum_{j,j' \in A} K_{j-j'} \langle a_j a_{j'}   \rangle e^{i[\phi_a(j) - \phi_a(j')]} - \nonumber \\
	&& \sum_{j,j' \in B} K_{j-j'}\langle a_j  a_{j'} \rangle e^{i[\phi_a(j) - \phi_a(j')]} ~.
\end{eqnarray}
We use then: 
\begin{eqnarray}
	 \sum_{j,j'=0}^{N-1} K_{j-j'} a_j a_{j'} e^{i[\phi_a(j) - \phi_a(j')]} =\nonumber \\
	  \frac{2}{N}\sum_{k=-\frac{N}{4}+1}^{\frac{N}{4}-1} \left|
	 	\sum_{r=0}^{N-1} a_r e^{2i\pi kr/N + i\phi_a(r)} \right|^2 
	 	\ge  0 ~.
\end{eqnarray}
We then observe that for $j,j'\in A$, $j-j'$ is an even integer, and therefore, from Eq.~\eqref{eq_expression_Kr_bipartite}, $K_{j-j'}=1 - 2 / N$ if $j=j'$ and $K_{j-j'} = -(2/N)(-1)^{(j-j')/2}$ if $j\neq j'$. Therefore, we have:
\begin{eqnarray}
	 \sum_{j,j'\in A} K_{j-j'} a_j a_{j'} e^{i[\phi_a(j) - \phi_a(j')]} = \nonumber \\
	  \sum_{r\in A} a_r^2 - \frac{2}{N}\left|\sum_{r \in A} (-1)^{r/2} a_r e^{i\phi_a(r)} \right|^2
	 \le  \sum_{r\in A} a_r^2 ~.
\end{eqnarray}
By the same argument, observing the $j-j'$ is an even integer for $j,j'\in B$, we have:
\begin{equation}
	\sum_{j,j'\in B} K_{j-j'} a_j a_{j'} e^{i[\phi_a(j) - \phi_a(j')]} \le \sum_{r\in B} a_r^2  ~.
\end{equation}
We therefore have: 
\begin{equation}
 2W_a \ge -\sum_{r=0}^{N-1} \langle a_r^2 \rangle ~.
\end{equation}
Combining these inequalities for $a \in \{X, Y, Z\}$, we conclude that:
\begin{equation}
	2(W_X + W_Y + W_Z) \ge -\sum_{r=0}^{N-1} \langle x_r^2+y_r^2+z_r^2 \rangle \ge -N ~,
\end{equation}
where in the last step we have used that the constraint \eqref{eq_local_constraint_LV} is obeyed by fully separable states. This achieves the proof that Eq.~\eqref{eq_witness_bipartite_appendix} is an entanglement witness. Combining with the observation made at the beginning of Sec. \ref{sec_bipartite_witnesses}, we conclude that a violation of such a witness directly implies bipartite entanglement, since its expression involves only cross-correlations between $A$ and $B$ subsystems.
\end{proof}

\section{Relation to previous entanglement criteria}
\label{app_previous_works}

Here we show how one can recover some previously known entanglement criteria as a consequence of the PSD condition introduced in Sec. \ref{sec_framework}.

\subsection{Recovering the covariance matrix criterion}
In this subsection, we show how the so-called covariance matrix criterion (CMC) \cite{gittsovichetal2010} for detecting entanglement in a multiqubit state is a consequence of our approach. In order to state the CMC within the notations of this paper, we consider three qubits (the $N$-qubit case requires a trivial generalization). Applying the CMC requires the knowledge of all one-body and two-body correlations for all pairs, and all Pauli matrices. We denote $C_i$ the vector $(C_i^X, C_i^Y, C_i^Z)$, and $C_{ij}$ the $3 \times 3$ matrix with entries $C_{ij}^{ab}$ [see Eq.~\eqref{eq_1body2body_correlators} for the definition of $C_i^a$ and $C_{ij}^{ab}$]. The CMC states that if the three-qubit state is fully separable, then there exist three real symmetric $3 \times 3$ matrices $\rho_i \succeq 0$, with ${\rm Tr}(\rho_i)=1$, such that:
\begin{equation}
\label{eq_CMC}
	\begin{pmatrix}
		\rho_1 & C_{12} & C_{13} \\
		C_{21} & \rho_2 & C_{23} \\
		C_{31} & C_{32} & \rho_3
	\end{pmatrix} \succeq \begin{pmatrix}
		C_1^T \\ C_2^T \\ C_3^T
	\end{pmatrix} \begin{pmatrix}
		C_1 & C_2 & C_3
	\end{pmatrix} ~.
\end{equation}
On the other hand, the first level of our hierarchy [see Eq.~\eqref{eq_Gamma_matrix}] implies that:
\begin{equation}
	\Gamma = \begin{pmatrix}
		1 & C_1 & C_2 & C_3 \\
		C_1^T & \sigma_1 & C_{12} & C_{13} \\
		C_2^T & C_{21} & \sigma_2 & C_{23} \\
		C_3^T & C_{31} & C_{32} & \sigma_3
	\end{pmatrix} \succeq 0 ~,
\end{equation}
with $\sigma_i = \begin{pmatrix}
	\langle x_i^2 \rangle & \langle x_i y_i \rangle & \langle x_i z_i \rangle \\
	\langle x_i y_i \rangle & \langle y_i^2 \rangle & \langle y_i z_i \rangle \\
	\langle x_i z_i \rangle & \langle y_i z_i \rangle & 1 - \langle x_i^2 \rangle - \langle y_i^2 \rangle
\end{pmatrix}$, where $\langle \dots \rangle$ denotes an average over the (classical) $p$ distribution defining a separable state (Section \ref{sec_framework}). Clearly, $\sigma_i \succeq 0$ is a symmetric matrix of unit trace. Furthermore, positivity of $\Gamma$ implies Eq.~\eqref{eq_CMC}. Notice that our approach is more general than the approach underlying the CMC, for at least three reasons: 1) we can naturally deal with missing entries in the correlation matrix (and actually leverage on it to strongly reduce the computational cost, see Appendix \ref{app_symmetries}); 2) the criterion of Eq.~\eqref{eq_Gamma_matrix} which implies the CMC is only the first level of a systematic hierarchy converging towards the separable set; 3) we provide a systematic approach to incorporate the knowledge of any correlation function, beyond one- and two-body considered in the CMC.

\subsection{Recovering the generalized spin-squeezing inequalities}
Here, we show how the generalized spin-squeezing inequalities derived in Ref.~\cite{tothetal2009} can be recovered within our approach. These inequalities are entanglement witnesses invariant under all permutations of the qubits, and may be defined in terms of first- and second-moments averaged over all permuatations: $\{m_a, C_{aa}; a \in \{x,y,z\}\}$ with $m_a=N^{-1}\sum_{i=1}^N C_i^a$ and $C_{aa} = [N(N-1)]^{-1}\sum_{i \neq j}C_{ij}^{aa}$. They consist of the following eight inequalities, valid for all fully-separable states [Eq.~(50) of Ref.~\cite{tothetal2009}]:
\begin{eqnarray}
	&C_{xx} + C_{yy} + C_{zz} \le 1 \nonumber \\
	&C_{xx} + C_{yy} + Nm_z^2 - (N-1) C_{zz} \le 1 \nonumber \\
	&C_{yy} + C_{zz} + Nm_x^2 - (N-1) C_{xx} \le 1 \nonumber \\
	&C_{zz} + C_{xx} + Nm_y^2 - (N-1) C_{yy} \le 1 \nonumber \\
	&C_{xx} + N(m_y^2 + m_z^2) - (N-1) (C_{yy} + C_{zz}) \le 1 \nonumber \\
	&C_{yy} + N(m_z^2 + m_y^2) - (N-1) (C_{zz} + C_{xx}) \le 1 \nonumber \\
	&C_{zz} + N(m_x^2 + m_y^2) - (N-1) (C_{xx} + C_{yy}) \le 1 \nonumber \\
	&N(m_x^2 + m_y^2 + m_z^2) - (N-1)(C_{xx} + C_{yy} + C_{zz}) \le  1 \nonumber
\end{eqnarray}
In order to prove these inequalities within our approach, we assume that there exists a probability distribution $p[\{x_i,y_i,z_i\}]$ reproducing the data. Following the symmetrization procedure described in Appendix \ref{app_symmetries}, without loss of generality we may choose the $p$ distribution invariant under all permutations. We then consider as monomials ${\bf u} = (1, x_1, \sum_{i=1}^N x_i)$, and build the corresponding correlation matrix $\Gamma_x = \langle {\bf u}^T {\bf u} \rangle$, where the average is over the $p$ distribution. Using the invariance of $p$ under permutations of the qubits, we obtain:
\begin{equation}
	\Gamma_x = \begin{pmatrix}
		1 & m_x & Nm_x \\
		\cdot & \langle x_1^2 \rangle & \langle x_1^2 \rangle + (N-1)C_{xx} \\
		\cdot & \cdot & N\langle x_1^2 \rangle + N(N-1) C_{xx} 
	\end{pmatrix} \succeq 0 ~.
\end{equation}
Positivity of $\Gamma_x$ implies that: $${\rm det}
	\begin{pmatrix} 
			1 & Nm_x \\
			Nm_x & N\langle x_1^2 \rangle + N(N-1) C_{xx}
	\end{pmatrix} \ge 0$$ and: $${\rm det}
	\begin{pmatrix} 
			\langle x_1^2 \rangle & \langle x_1^2 \rangle + (N-1)C_{xx} \\
			\langle x_1^2 \rangle + (N-1)C_{xx} & N\langle x_1^2 \rangle + N(N-1) C_{xx} 
	\end{pmatrix} \ge 0$$
From these inequalities, we obtain the conditions:
\begin{eqnarray}
	&C_{xx} \le \langle x_1^2\rangle \\
	&Nm_x^2 - (N-1) C_{xx} \le  \langle x_1^2 \rangle
\end{eqnarray}
For the same reason, we also have:
\begin{eqnarray}
	&C_{yy} \le \langle y_1^2\rangle \\
	&Nm_y^2 - (N-1) C_{yy} \le  \langle y_1^2 \rangle \\
	&C_{zz} \le \langle z_1^2\rangle \\
	&Nm_z^2 - (N-1) C_{zz} \le  \langle z_1^2 \rangle
\end{eqnarray}
Combining these inequalities, and using the property $\langle x_1^2 + y_1^2 + z_1^2 \rangle \le 1$, we recover the eight inequalities of Ref.~\cite{tothetal2009}. The approach presented in this paper is however much more general. Notice also that a straightforward extension to qudits allows one to recover the results of Ref.~\cite{vitaglianoetal2011}.

\end{document}